\documentclass[twocolumn,floatfix]{revtex4}
\usepackage{color}
\usepackage{epsfig}
\usepackage{graphicx}
\usepackage{amsmath}

\newcommand {\be}  {
\begin{equation}
}
\newcommand {\ee}  {
\end{equation}
}
\newcommand {\bea} {
\begin{eqnarray}
}
\newcommand {\eea} {
\end{eqnarray}
}

\newcommand {\ma} {
\mathcal A
}

\newcommand {\mb} {
\mathcal B
}

\newcommand {\mc} {
 \mathcal C
}

\newcommand {\m} {
 \mathcal 
}

\newcommand{\boldnabla}{\text{\boldmath$\nabla$}}

\begin{document}

%\title{jhkjhkjh}

%\author{C. H\"uter$^1$}
%\author{G. Boussinot$^2$}
%\author{E. A. Brener$^2$}
%\author{D. E. Temkin}
%\affiliation{$^1$Computational Materials Design Department, 
%Max-Planck Institut f\"ur Eisenforschung, D-40074 D\"usseldorf, Germany}
%\affiliation{$^2$Institut f\"ur Festk\"orperforschung, 
%Forschungszentrum J\"ulich, D-52425 J\"ulich, Germany}

%\pacs{45.70.Qj, 68.08.-p, 81.30.Fb }

%\date{\today}

\title{Interface kinetics in phase field models: isothermal transformations in binary alloys and steps dynamics in molecular-beam-epitaxy}
\author{ G. Boussinot$^{1,2}$ and Efim A. Brener$^2$
%Efim A. Brener$^1$ and G. Boussinot$^{1,2}$
}
\affiliation{$^1$Computational Materials Design Department, 
Max-Planck Institut f\"ur Eisenforschung, D-40237 D\"usseldorf, Germany\\
$^2$Peter Gr{\"u}nberg Institut, Forschungszentrum J{\"u}lich, D-52425 J{\"u}lich, Germany 
}

\begin{abstract}

%We present a unified description of interface kinetic effects in phase field models for growth processes coupled with diffusion in the bulk. We focus on isothermal transformations in binary alloys and steps dynamics in molecular-beam-epitaxy. 
We present a unified description of interface kinetic effects in phase field models for isothermal transformations in binary alloys and steps dynamics in molecular-beam-epitaxy. 
The phase field equations of motion incorporate a kinetic cross-coupling between the phase field and the concentration field. This cross coupling generalizes the phenomenology of kinetic effects and was omitted until recently in classical phase field models. 
We derive general expressions (independent of the details of the phase field model) for the kinetic coefficients within the corresponding macroscopic approach using a physically motivated reduction procedure. The latter is equivalent to the so-called thin interface limit but is technically simpler. It involves the calculation of the effective dissipation that can be ascribed to the interface in the phase field model. We discuss in details the possibility of a non positive definite matrix of kinetic coefficients, i.e. a negative effective interface dissipation, although being in the range of stability of the underlying phase field model. 
Numerically, we study the step-bunching instability in molecular-beam-epitaxy due to the Ehrlich-Schwoebel effect, present in our model due to the cross-coupling. Using the reduction procedure we compare the results of the phase field simulations with the analytical predictions of the macroscopic approach.
% This comparison includes some range of parameters where the step's dissipation is effectively negative, i.e. when the macroscopic Onsager matrix of kinetic coefficients is not positive definite. 

%We present a calculation of the energy and the dimensions of the critical fluctuations required for the appearance of an elongated crystal perpendicular to a substrate with a condition of coherence at the interface between them. We present our analysis in the grand canonical ensemble where the number of atoms is not fixed. This can be interpreted as a theory for heterogeneous nucleation of a solid from the melt with coherence condition at the wall. We relate our work to the commonly admitted theories of island formation (CITE), and in particular, we show that our theory relaxes the condition of small slope assumed in these theories.

\end{abstract}

\maketitle

 \section{Introduction}
 
Phase field models have become a commonly used numerical tool in a wide range of pattern formation processes such as solidification \cite{book}, solid-solid transformations \cite{chen} or fluid mechanics \cite{anderson}, and also in other fields of materials science, physics, biophysics and engineering. Designed in the spirit of the Ginzburg-Landau theory for phase transitions, they avoid a direct tracking of the boundaries between different phases by the introduction of continuous fields varying smoothly across these boundaries or interfaces. One then refers to diffuse interface models and the interface width $W$ is a key parameter of these models that has to be handled with care.
The equations of motion in phase field models are solved everywhere in the simulation domain and replace the set of bulk equations and interfaces boundary conditions of the macroscopic approach where the interfaces are assumed to be infinitely sharp. 
%The interface width should then be much smaller than any macroscopic length scale that arises from the bulk equations.

In crystal growth, linear interface kinetics describe small deviations from local equilibrium boundary conditions at the interfaces. While, in many cases, kinetic effects are subdominant in comparison with the large dissipation in the bulk, they are crucial ingredients for the description of  important physical phenomena such as the solute trapping effect in binary alloys or the Ehrlich-Schwoebel effect in molecular-beam-epitaxy. The kinetic coefficients  give the proportionality between driving forces and fluxes in the frame of Onsager theory of linear out-of-equilibrium thermodynamics. In the bulk, the Onsager theory links the diffusion flux and the gradient of chemical potential. The situation is more complicated at a moving interface. Indeed, for example in the case of a binary A-B alloy, the growing phase may incorporate different amounts of B atoms for a given growth velocity. In other words, the concentration of B atoms on the two sides of the interface are independent variables. Therefore there exist in this case two independent fluxes of atoms through the interface, i.e. the total number of atoms and the number of B atoms. The linear relations between the driving forces (in this case the difference of chemical potentials between the two sides of the interface) and these fluxes are provided by a $2\times 2$ Onsager matrix of kinetic coefficients. Due to Onsager symmetry this matrix contains three independent elements. 
In the steps dynamics problem in molecular-beam-epitaxy (MBE), there exist also two independent fluxes through the steps and thus three independent kinetic coefficients should be considered. Since the pioneering work by Burton, Frank and Cabrera \cite{bcf}, huge theoretical efforts have been devoted to the description of steps dynamics on vicinal surfaces on which the presence of atomic steps is inherent. For a recent review on the macroscopic approach in MBE see, for example, Ref. \cite{misbah} and references therein. One of the feature of step kinetics is the unequal probabilities of attachment from lower and upper terraces, i.e. the Ehrlich-Schwoebel (ES) effect \cite{ehrlich, schwoebel}. It is responsible for instabilities \cite{misbah} such as step-bunching \cite{schwoebel} or meandering \cite{bales}.

In classical phase field models for growth processes coupled with diffusion in the bulk (based on the model C within the classification of Hohenberg and Halperin \cite{hohenberg}), only two independent kinetic coefficients were introduced. 
%Apart from the metallurgical problems like solidification of pure materials CITE or solidification in binary alloys CITE, one should mention for the steps dynamics the work on spiral growth from a screw dislocation \cite{karma_spiral} or on the dynamics of a localized trains of steps \cite{liu}.
The variational formulation of these models, that links the time derivatives of the fields to the functional derivatives of the free energy $G$ with respect to them, is diagonal. This means that the time derivative of the phase field $\phi$ is $\tau \dot \phi = -\delta G/\delta \phi$ and the time derivative of the concentration field $C$ is  $\dot C = D \boldnabla^2 \delta G/\delta C$. Two velocity scales describing the interface kinetics are then built using the interface width $W$, i.e. $W/\tau$ and $D/W$.

Here we present phase field models for isothermal transformations in binary alloys and for steps dynamics in MBE that incorporate the third kinetic coefficient due to cross terms introduced in the equations of motion. The magnitude of these cross effects is constrained by the positive definiteness of the dissipation in the system. These cross effects were first described in a phase field model for isothermal transformations in binary alloys recently  published as a Rapid Communication \cite{brener_boussinot}. They were also introduced in Ref. \cite{fang} to recover the thermodynamical consistency of the anti-trapping model \cite{antitrapping}. 

%A question arises concerning the relation between a given phase field model and the  corresponding macroscopic description. 
For a given phase field model, a question arises concerning its relation to the macroscopic description.
Therefore in addition to the presentation of the phase field models, we propose a procedure, involving the calculation of the interface dissipation function, that links the parameters of the model to the kinetic boundary conditions in the macroscopic approach. It is done in a general way, i.e. independent of the details of the phase field model.
This physically motivated procedure is equivalent to the more formal asymptotic matching within the thin interface limit \cite{karma_rappel} but it is, in our opinion, technically simpler.  The well-known results of the thin interface limit for binary alloys are then derived in very concise terms. 
We stress that the domain of stability of the phase field model (positive definiteness of the dissipation) is wider than the domain of stability of the corresponding macroscopic approach (positive definiteness of the matrix of macroscopic kinetic coefficients). In other words, in some range of parameters of a stable phase field model, the interface dissipation of the corresponding macroscopic description may be negative. While this fact is known \cite{karma_rappel, elder} ("negative growth kinetic coefficient"), we discuss this non-trivial issue in details.  
We also give the results of the reduction procedure for the steps dynamics in MBE. It turns out that the newly introduced cross coupling in the phase field equations of motion is responsible for the ES effect. Note that the latter was absent in classical diagonal phase field models for the dynamics of a localized train of steps \cite{liu} and for spiral growth from a screw dislocation \cite{karma_spiral}. 
Note also that the ES effect was described in a phase field model for step dynamics using a different philosophy where each terrace on the vicinal surface possesses a separate concentration field \cite{pierre_louis}. 
%Our reduction procedure allows us to compare simulations results to the analytical results obtained within the macroscopic approach. We perform this comparison for the step-bunching instability on a vicinal surface, stressing results where the matrix of macroscopic kinetic coefficients is non-positive definite.   

In the first part of this article, we present the phase field model, the general reduction procedure and explicit examples of the latter for the case of phase transformations in binary alloys. In the second part, we study steps dynamics in MBE, presenting the phase field model, the link with the macroscopic description and finally simulations results of step-bunching instability. We compare these numerical results with analytics within the macroscopic approach, stressing the case where the matrix of macroscopic kinetic coefficients is non-positive definite.

\section{Phase field model of isothermal phase transformations in binary alloys} \label{alloy}

In phase field models of phase transformations in binary alloys, the motion of the boundary between two different phases involves a scalar order parameter $\phi$ that discriminates the phases and is a non conserved field. To describe diffusion processes, one additionally has a concentration field $C$ which is conserved. We introduce a free energy functional in a standard dimensionless form: 
\bea \label{free_energy}
G[C,\phi]=\int dV \left\{H \left[\frac{(W\boldnabla\phi)^2}{2}+f(\phi) \right]  +g(C,\phi)\right\} .
\eea
The phase field $\phi$ is constant in the bulk of each phase corresponding to the values that are minimizing the double well potential $f(\phi)$, and which are usually integer values, for example 0 and 1. The phase field then varies from $\phi=0$ in phase 0 to $\phi=1$ in phase 1 across the interface of width $W$, i.e. $W|\boldnabla \phi| \sim 1$. The energetic cost of the interface is described by $H$ which is usually a large parameter. $g(C,\phi)$ describes a thermodynamic coupling between the phase field and the concentration field. The functions $g(C,\phi=0)=g_0(C)$ and $g(C,\phi=1)=g_1(C)$ should then describe the free energy density $g_0$ and $g_1$ of phase 0 and phase 1 respectively as a function of $C$ (we omit the temperature variable since we discuss isothermal transformations). The equilibrium one-dimensional distributions (coordinate $x$) are $\phi=\phi_{eq}(x)$ that verifies $[W\phi_{eq}'(x)]^2 = 2f[\phi_{eq}(x)]$ and $C=C_{eq}(x)$ that verifies $\frac{\partial g}{\partial C}[C_{eq}(x), \phi_{eq}(x)] = g_1'(C_1^{eq}) = g_0'(C_0^{eq}) = [g_1(C_1^{eq}) - g_0(C_0^{eq})]/(C_1^{eq} - C_0^{eq})$ where $C_1^{eq}$ ($C_0^{eq}$) is the equilibrium concentration in phase 1 (phase 0).

\subsection{Equations of motion}

On one hand, $\dot \phi$, which is non vanishing only in the interface region and is proportional to the normal velocity of the interface, represents the amount of matter that undergoes the phase transformation per unit time. It is therefore a 'flux' of atoms through the interface that is linearly related to some driving forces, in the frame of Onsager theory of out-of-equilibrium thermodynamics, and that accounts for interfacial kinetic effects. On the other hand $C$ is a conserved field and obeys the continuity equation 
\be \label{continuity_eq}
\dot C = - \boldnabla \cdot \bf J
\ee
 where $\bf J$ is the diffusional flux. This equation holds in the bulk and in the interface region. Therefore $\bf J$ plays a two-fold role: it describes the diffusion in the bulk and it is the second flux that accounts for kinetic effects at the interface.

The driving forces, to which $\dot \phi$ and $\bf J$ are linearly related in the framework of Onsager theory, are derivatives of the free energy functional $G$ with respect to the fields $\phi$ and $C$ (variational formulation). The driving force conjugated to $\dot \phi$ is $-\delta G/\delta \phi$ and the driving force conjugated to $\bf J$ is $-\boldnabla \delta G/\delta C$. The phase field equations of motion give the linear relations between the driving forces and the fluxes with the mean of a $2 \times 2$ symmetric matrix and we choose in the following to express driving forces in terms of fluxes (in Appendix \ref{other_pf} we present the equations of motion for a matrix giving fluxes in terms of driving forces). The equation giving $-\delta G/\delta \phi$ is scalar and the equation giving $-\boldnabla \delta G/\delta C$ is vectorial. Thus, the diagonal elements of the Onsager matrix, giving the proportionality between conjugate quantities, are scalar and the non diagonal element or cross term is vectorial. Moreover the equation giving $-\delta G/\delta \phi$ has to vanish in the bulk even though $\bf J$ does not. Therefore the cross term is very naturally written proportional to $W\boldnabla \phi$ which is a vector perpendicular to the interface that has a vanishing norm in the bulk and a norm of order unity within the interface. 
The phase field equations are thus written:  
\bea \label{phase_field}
 -\frac{\delta G}{\delta \phi} \; = \; \tau(\phi) \;\dot \phi \; +\;   [ M(\phi) W \boldnabla \phi ] \cdot {\bf J} \;,\\
-\boldnabla \frac{\delta G}{\delta C} \; = \; [ M(\phi) W \boldnabla \phi ] \; \dot \phi \; + \; \frac{{\bf J}}{D(\phi)} \label{diff_flux} \;.
 \eea
 The diagonal terms are parametrized by the time scale $\tau(\phi)$ and the diffusion coefficient $D(\phi)$ \cite{comment}. The cross terms are parametrized by the inverse velocity scale $M(\phi)$.
  
Note that the use of $\boldnabla \phi$ to introduce a vectorial quantity out of the scalar field $\phi$ was initiated in the anti-trapping model \cite{antitrapping}. However, this model does not obey Onsager symmetry since it introduces the cross term in Eq. (\ref{diff_flux}) and not in Eq. (\ref{phase_field}). This was noted only very recently \cite{brener_temkin} and cross terms were then introduced in a proper way in Refs. \cite{brener_boussinot} and \cite{fang}.

Note also that linear out-of-equilibrium thermodynamics correspond to small deviations of the fluxes $\dot \phi$ and $\bf J$ from 0. Therefore, the quantity $[M(\phi)W\boldnabla \phi]$ may be introduced in Eqs. (\ref{phase_field}) and (\ref{diff_flux}) through its equilibrium distribution $[M(\phi_{eq})W\boldnabla \phi_{eq}]$ where $\phi_{eq}$ verifies $\delta G/\delta \phi = 0$. However it is more computationally convenient to calculate the gradient of the existing field $\phi$ than having the equilibrium distribution $\phi_{eq}$ as an input. Close to equilibrium the two possibilities are equivalent.
\\

{\it Stability and dissipation.}
To ensure the thermodynamical stability of the phase field model, the diagonal terms have to be positive:
\bea
\tau(\phi) > 0, \text{ and }
D(\phi) > 0 . \label{stability_diag}
\eea
In addition, the determinant 
\be\label{determinant}
\Delta_{PF} = 1 - \frac{[ M(\phi)W \boldnabla \phi]^2 D(\phi)}{\tau(\phi)}
\ee
 must also be positive, leading to some restriction on the absolute value of $M(\phi)$. Close to equilibrium this restriction reads:
%\be\label{stability}
%M^2 < \frac{\tau}{\text{max} [D(\phi)(W\boldnabla \phi)^2]} \;.
%\ee
\be\label{stability}
M^2(\phi_{eq}) < \frac{\tau(\phi_{eq})}{ D(\phi_{eq})(W \boldnabla\phi_{eq})^2} \;.
\ee
The inequalities (\ref{stability_diag}) and (\ref{stability}) ensure that the dissipation 
\bea \label{dissipation_pf}
R &=& \frac{1}{2} \int_V dV \left[ - \dot \phi \frac{\delta G}{\delta \phi} - {\bf J} \cdot \boldnabla \frac{\delta G}{\delta C}\right] \\
&=& \frac{1}{2} \int_V dV \left[ \tau(\phi)  \left[\dot \phi\right]^2 + \frac{{\bf J}^2}{D(\phi)} + 2M(\phi)W\dot \phi \boldnabla \phi\cdot {\bf J} \right] \nonumber
\eea
is positive whatever $\phi$ and $C$.

\subsection{Reduction to the kinetic boundary conditions in the macroscopic approach} \label{reduction_alloy}

In the macroscopic description, the interface is a sharp boundary (zero thickness) between domains where the bulk equations hold. The free boundary problem then requires some conditions at the interface. First, one has a mass conservation equation. Second, one has to prescribe the concentration on both sides of the interface.
Without kinetic effects, the concentrations at the interface are the equilibrium ones (possibly incorporating a Gibbs-Thomson correction). When kinetic effects are present, the concentrations deviate from equilibrium ones. Within the Onsager approach of out-of-equilibrium thermodynamics, these deviations are representing driving forces that are linearly related to some fluxes through the interface. Driving forces and fluxes should be chosen appropriately in order to have couples of conjugated quantities.  

In the case of binary alloys or in the case of steps dynamics in MBE (that are closely related formally), two driving forces and two fluxes are required to describe interface kinetics. We therefore have three independent kinetic coefficients that are elements of a $2\times2$ symmetric Onsager matrix. The choice of the two couples of conjugate quantities (the basis) is completely arbitrary and each choice is valid. However, each problem has its own commonly used basis. We will present in the next section about steps dynamics in MBE the link between the kinetic coefficients whether using the commonly used basis in binary alloy problems or the commonly used basis in MBE. 

\subsubsection{Kinetic boundary conditions}

In binary A-B alloys problems, one has, in dimensionless form, the chemical potential of A atoms in phase $i$ ($i=0,1$), $\mu_A^{(i)}(C) = g_i(C) - Cg_i'(C)$, and the chemical potential of B atoms in phase $i$, $\mu_B^{(i)}(C)=g_i(C)+(1-C)g_i'(C)$. $C$ is the concentration of B atoms and $g_i(C)$ is the dimensionless free energy density as a function of $C$ of phase $i$ as mentioned before. One usually uses instead the grand potential $\mu_A^{(i)}(C)$ and the diffusion chemical potential $\mu^{(i)}(C) = \mu_B^{(i)}(C)-\mu_A^{(i)}(C) = g_i'(C)$.
For an interface between phase 1 and phase 0, one then considers the driving forces 
\bea
\delta \mu_A (C_1,C_0)= \mu_A^{(0)}(C_0) - \mu_A^{(1)}(C_1) , \nonumber\\
\delta \mu (C_1,C_0)= \mu^{(0)}(C_0) - \mu^{(1)}(C_1) , \nonumber
\eea
where $C_i$ is the concentration in phase $i$ at the interface.
At equilibrium, we have $\delta \mu_A(C^{eq}_1,C^{eq}_0)=\delta \mu (C^{eq}_1,C^{eq}_0)= 0$ with $C_i^{eq}$ the equilibrium concentration in phase $i$.
Near equilibrium, we have
\bea
\delta \mu_A & \approx& C_1^{eq}(C_1-C_1^{eq}) g''_1(C_1^{eq})  - C_0^{eq}(C_0-C_0^{eq}) g''_0(C_0^{eq}),  \nonumber \\
\delta \mu  &\approx& (C_0-C_0^{eq}) g''_0(C_0^{eq}) - (C_1-C_1^{eq}) g''_1(C_1^{eq}),  \label{delta_mu_exp}
\eea
where $g_i''(C)$ is the second derivative of $g_i(C)$ with respect to $C$.
One then writes the linear relations
\bea
\delta \mu_A = \bar \ma V + \bar \mb J_B, \label{deltamua}\\
\delta \mu = \bar \mb V + \bar \mc J_B, \label{deltamu}
\eea
where $V$ is the flux conjugated to $\delta \mu_A$ and $J_B$ the flux conjugated to $\delta \mu$. $V$ represents the total flux of atoms through the interface (atomic volume times the number of atoms A and B that are undergoing the phase tranformation per unit time and per unit area) and is actually the normal velocity of the interface. $J_B$ is the flux through the interface of B atoms only. 
$\bar \ma, \bar \mb$ and $\bar \mc$ are the three independent Onsager kinetic coefficients. 
The Onsager matrix is definite positive if the "growth kinetic coefficient" is positive $\bar \ma>0$, the "diffusional resistance" of the interface (in analogy to the Kapitza resistance in the thermal problem) is positive $\bar \mc>0$ and the determinant is positive $\bar \ma\bar \mc-\bar \mb^2>0$. The positive definiteness ensures that the interface dissipation 
\be\label{rint}
R_{int}=(\delta \mu_A V + \delta  \mu J_B)/2=\bar \ma V^2/2 + \bar \mc J_B^2/2 + \bar \mb V J_B
\ee
 is positive whatever $V$ and $J_B$.
 The normal gradients of concentration at the interface, i.e the diffusion fluxes through the interface, are related to $V$ and $J_B$ using the following mass conservation equations \cite{temkin}:
 \bea
 - D_1 \boldnabla C|_1 \cdot {\bf n} = V C_1 - J_B,  \label{flux1} \\
 - D_0 \boldnabla C|_0 \cdot {\bf n} = V C_0 - J_B \;. \label{flux2}
 \eea
$D_1$ ($D_0$) is the diffusion coefficient in phase 1 (phase 0), $\boldnabla C|_1$ ($\boldnabla C|_0$) is the gradient of concentration at the interface on the side of phase 1 (phase 0) and $\bf n$ is the normal to the interface. 
%Subtracting the last two equations allows one to recover the usual mass conservation equation used in free boundary problems of diffusional growth.

%This macroscopic description does not presuppose any detail of the interface region and any thermodynamically consistent phase field model of the interface may be reduced to this macroscopic description near equilibrium. This means that one may find the coefficients $\bar \ma, \bar \mb$ and $\bar \mc$ as functions of the phase field parameters. 

\subsubsection{Link between boundary conditions and phase field parameters: reduction procedure.} \label{reduction_alloy_sub}

We now present the procedure to determine the correspondence between the coefficients $\bar \ma, \bar \mb$ and $\bar \mc$ and the parameters of the phase field model. Let us consider the coordinate $x$ of a one-dimensional infinite system with an interface centered at $x=0$. The interface connects phase 1 ($\phi(-\infty) = 1$) and phase 0 ($\phi(+\infty) = 0$). The total dissipation in this system, see Eq. (\ref{dissipation_pf}), is
\bea
R &=& \frac{1}{2} \int_{-\infty}^{-\delta} dx \frac{J^2(x)}{D_1} + \frac{1}{2} \int_{\delta}^{\infty} dx \frac{J^2(x)}{D_0} \nonumber \\
&& + \frac{1}{2} \int_{-\delta}^\delta dx \left[ \tau(\phi) \left[\dot \phi \right]^2 + \frac{J^2(x)}{D(\phi)} + 2M(\phi)W\dot \phi \phi'(x) J(x) \right], \nonumber
\eea  
where $\delta \sim W$ is such that $\phi(x<-\delta) \approx 1$ and $\phi(x>\delta) \approx 0$, and where $D_1 = D(\phi=1)$ and $D_0 = D(\phi=0)$. In the bulk ($|x|>\delta$) where $\dot \phi=0$ and $\phi'=0$, only the diffusion flux $J(x)={\bf J}$ contributes to the dissipation and the macroscopic length scale that characterizes its variations is much larger than $\delta \sim W$. Within the macroscopic approach with an infinitely sharp interface, the dissipation in the same system is expressed through the diffusional flux in phase 1 and in phase 0 (which are corresponding to $J(x)$ for $x<-\delta$ and $x>\delta$ respectively) as:
\be
\frac{1}{2} \int_{-\infty}^0 dx \frac{J^2(x)}{D_1} + \frac{1}{2} \int_{0}^\infty dx \frac{J^2(x)}{D_0} + R_{int} \;. \nonumber
\ee  
In order for the latter dissipation function to be equal to $R$, and taking into account that $J(-\delta>x>0) \approx J(-\delta)= J_1$ and $J(0<x<\delta) \approx J(\delta)= J_0$ due to the small variations of $J(x)$ in the bulk, one may write the dissipation ascribed to the interface within the phase field model as
\bea\label{rint_pf1}
R_{int} &=& \frac{1}{2} \int_{-\delta}^\delta dx \left[ \tau(\phi) \left[ \dot \phi \right]^2 + \frac{J^2(x)}{D(\phi)} + 2M(\phi)W\dot \phi \phi'(x) J(x) \right] \nonumber  \\
&&- \frac{1}{2} \int_{-\delta}^0 dx \frac{J_1^2}{D_1} - \frac{1}{2} \int_{0}^\delta dx \frac{J_0^2}{D_0} \;.
\eea
In order to identify Eq. (\ref{rint_pf1}) with Eq. (\ref{rint}), one should express the fluxes $\dot \phi$ and $J(x)$ in terms of $V$ and $J_B$. This is done using a quasi-steady approximation that assumes large gradients of $\phi$ and $C$ across the interface compared with bulk ones. This gives for $\dot \phi$:
\be \label{phi_dot}
\dot \phi \approx -V \phi'(x) \;.
\ee
For the concentration field, the quasi-steady approximation $\dot C \approx -VC'(x)$ allows to integrate the continuity equation $\dot C=-J'(x)$. Then, choosing $-J_B$ as an integration constant yields:
\be \label{j(x)}
J(x) \approx V C(x) - J_B \;,
\ee
which corresponds to Eqs. (\ref{flux1}) and (\ref{flux2}) near the interface.
Close to equilibrium, i.e. for linear kinetic effects, we have $V \phi'(x) \approx V\phi'_{eq}(x)$ and $VC(x) \approx VC_{eq}(x)$ where $\phi_{eq}(x)$ and $C_{eq}(x)$ are the equilibrium distributions of $\phi$ and $C$. $R_{int}$ in Eq. (\ref{rint_pf1}) may therefore be written
%\bea \label{rint_pf2}
%R_{int} &=& \frac{1}{2} \int_{-\delta}^\delta dx \Big[ \tau [\phi'_{eq}(x)]^2V^2 + \frac{\big(V C_{eq}(x) - J_B \big)^2}{D(\phi)} \nonumber \\
%&&- 2MW  [\phi_{eq}'(x)]^2 V \big(V C_{eq}(x) - J_B \big) \Big] \nonumber  \\
%&&- \frac{1}{2} \int_{-\delta}^0 dx \frac{J_1^2}{D_1} - \frac{1}{2} \int_{-\delta}^0 dx \frac{J_0^2}{D_0} \;,
%\eea
\bea \label{rint_pf2}
R_{int} &=& \frac{1}{2} \int_{-\delta}^\delta dx \Big[ \tau(\phi_{eq}) [\phi'_{eq}(x)]^2V^2  \nonumber \\
&&- 2M(\phi_{eq})W  [\phi_{eq}'(x)]^2 V \big(V C_{eq}(x) - J_B \big) \Big] \nonumber  \\
&&+ \frac{1}{2} \int_{-\delta}^\delta dx \left[ \frac{\big(V C_{eq}(x) - J_B \big)^2}{D(\phi_{eq})} -  \frac{J_1^2}{2D_1} -  \frac{J_0^2}{2D_0} \right] \;, \nonumber \\
\eea
where $J_1 \approx VC_1^{eq} - J_B$ and $J_0 \approx VC_0^{eq} - J_B$ with $C_1^{eq}$ and $C_0^{eq}$ the equilibrium concentrations of phase 1 and phase 0 respectively. The range of integration $\delta$ is chosen such that $\phi'_{eq}(|x|>\delta) \approx 0$, $C_{eq}(x<-\delta) \approx C_1^{eq}$ and $C_{eq}(x>\delta) \approx C_0^{eq}$. On one hand, the integrand of the first integral thus vanishes for $|x|>\delta$. On the other hand, in the second integral, the integrand for $x<-\delta$ is the opposite of the integrand for $x>\delta$.
Therefore, the integrations in Eq. (\ref{rint_pf2}) may be performed from $-\infty$ to $+\infty$ leaving $R_{int}$ unchanged and independent of $\delta$. 
Identifying with Eq. (\ref{rint}) then yields
\bea \label{bara}
\bar \ma &=&  \int_{-\infty}^\infty dx \; \tau(\phi_{eq}) [\phi_{eq}'(x)]^2 \nonumber \\
&&-2  \int_{-\infty}^\infty dx \;M(\phi_{eq}) W[\phi_{eq}'(x)]^2 C_{eq}(x) \nonumber \\
&& +\int_{-\infty}^\infty dx \; \left[\frac{C^2_{eq}(x)}{D(\phi_{eq})} - \frac{(C_1^{eq})^2}{2D_1} - \frac{(C_0^{eq})^2}{2D_0} \right]  \;,
\eea
\bea\label{barb}
 \bar \mb &=&  \int_{-\infty}^\infty dx \; M(\phi_{eq})W [\phi_{eq}'(x)]^2 \nonumber \\
 &&- \int_{-\infty}^\infty dx \; \left[\frac{C_{eq}(x)}{D(\phi_{eq})} - \frac{C_1^{eq}}{2D_1} - \frac{C_0^{eq}}{2D_0} \right] \;,
\eea
\bea\label{barc}
\bar \mc = \int_{-\infty}^\infty dx \left[\frac{1}{D(\phi_{eq})} - \frac{1}{2D_1} - \frac{1}{2D_0} \right]  \;.
\eea  
This physically motivated and rather technically simple reduction procedure for deriving $\bar \ma, \bar \mb$ and $\bar \mc$ basically corresponds to the idea underlying the asymptotic matching in the thin-interface limit \cite{karma_rappel}. Although no doubt exists concerning the ability of the latter to reproduce the results given by Eqs. (\ref{bara}), (\ref{barb}) and (\ref{barc}), we did not find in the literature such a presentation. This set of equations is one of the main results in this article. Indeed, it provides the link in very general terms between the parameters entering the equations of motion Eqs. (\ref{phase_field}) and (\ref{diff_flux}) and the kinetic boundary conditions (\ref{deltamua}) and (\ref{deltamu}) at the interface. The specification of $\tau(\phi)$, $M(\phi)$, $D(\phi)$, $\phi_{eq}(x)$ and $C_{eq}(x)$ through the details of the phase field model then allows to have explicit expressions for $\bar \ma, \bar \mb$ and $\bar \mc$. We give in the following the explicit results of this reduction procedure for constant $\tau$ and $M$ in two cases: for a constant diffusion coefficient and for the one-sided model (where the diffusion is neglected in the growing phase). The general equations (\ref{bara}), (\ref{barb}) and (\ref{barc}) will also be used in order to derive the kinetic boundary conditions corresponding to the phase field model for step dynamics in MBE presented in the next section.

An alternative way to derive the coefficients $\bar \ma, \bar \mb$ and $\bar \mc$, that is followed in our previous article \cite{brener_boussinot}, consists in integrating the equations of motion (\ref{phase_field}) and (\ref{diff_flux}) across the interface. The chemical potentials are then calculated at a distance of order $W$ away from the center of the interface. This corresponds to the omission of the subtraction of the last two terms in Eq.(\ref{rint_pf1}). The kinetic coefficients $\bar \ma, \bar \mb$ and $\bar \mc$ then depend on the range of integration. In the next paragraph we discuss this issue with an explicit example. However, the present description with the subtraction of the bulk dissipation in the interface region corresponds to the asymptotic matching in the thin interface limit and is necessary in order to properly derive the macroscopic kinetic boundary conditions.
%The integration of the right-hand-sides gives expressions in terms of $V$ and $J_B$ when using the approximations in Eqs. (\ref{phi_dot}) and (\ref{j(x)}).
%The integration of the driving forces (left-hand-sides) provides differences of chemical potentials calculated at a distance of order $W$ away from the center of the interface $x=0$. $\delta \mu$ and $\delta \mu_A$ are then obtained using an extrapolation of the concentration field at $x=0$ that involves the normal gradient of concentration. This gives additional contributions to $\bar \ma, \bar \mb$ and $\bar \mc$ calculated using Eqs. (\ref{flux1}) and (\ref{flux2}). These additional contributions correspond to the subtraction of the last two terms on the r-h-s of Eq.(\ref{rint_pf1}). However, in Ref. \cite{brener_boussinot}, no extrapolation is performed and the coefficients $\bar \ma, \bar \mb$ and $\bar \mc$ are obtained through the chemical potentials calculated at a distance $\sim W$ away from the center of the interface. Equivalently, it means that the subtraction of the last two terms is omitted in Eq.(\ref{rint_pf1}).
%The kinetic coefficients $\bar \ma, \bar \mb$ and $\bar \mc$ then depend on the range of integration. In the next paragraph we will compare in this respect the expressions for $\bar \ma, \bar \mb$ and $\bar \mc$ obtained using Eqs. (\ref{bara}), (\ref{barb}) and (\ref{barc}) with the expressions for $\bar \ma, \bar \mb$ and $\bar \mc$ given in Ref. \cite{brener_boussinot} in the case of a constant diffusion coefficient.
% that are differing in this respect.

The Eq. (\ref{rint_pf1}) or (\ref{rint_pf2}) for $R_{int}$ involves integrals over a range of order $W$. The reduction procedure presented here may thus be used for a curved interface, $x$ representing the normal direction, as long as its curvature is much smaller than $1/W$. Then $\delta \mu_A$ is corrected by the Gibbs-Thomson effect which is proportional to the interface energy and may be obtained by the integration of the laplacian of $\phi$ in Eq. (\ref{phase_field}). We do not discuss surface diffusion and stretching effects \cite{antitrapping} at a curved interface that are generically smaller than kinetic effects in the macroscopic limit. The interface energy may depend on the orientation of the interface and such a dependence should then be introduced in the phase field model through an orientation dependence of $W$.
Moreover, the kinetic properties of  the interface may also depend on its orientation, and the phase field parameters $\tau(\phi), D(\phi)$ and $M(\phi)$ may then exhibit such a dependence. However, we do not discuss those issues here.

%One should note that an alternative way, that was followed in Ref. \cite{brener_boussinot}, for this reduction procedure consists in integrating Eqs. (\ref{phase_field}) and (\ref{diff_flux}) across the interface in order to provide an equation for $\delta \mu_A$ and an equation for $\delta \mu$. It then allows to recover the Gibbs-Thomson correction in the equation for $\delta \mu_A$ by the integration of the laplacian of $\phi$ in Eq. (\ref{phase_field}). Moreover we do not discuss the surface diffusion and stretching effects \cite{antitrapping} that are negligible with respect to kinetic effects in the macroscopic limit.

\subsubsection {Positiveness and non-positiveness of the Onsager matrix} \label{reduction_alloy_positiveness}

 It is clear from the expressions for $\bar \ma$, $\bar \mb$ and $\bar \mc$ given by Eqs. (\ref{bara}), (\ref{barb}) and (\ref{barc}) that, even if the phase field model is perfectly stable, the subtraction in Eq. (\ref{rint_pf1}) or (\ref{rint_pf2}) does not guarantee the positiveness of the Onsager matrix of kinetic coefficients in the corresponding macroscopic description. In other words, fulfilling the conditions $\tau(\phi)>0$, $D(\phi)>0$ and the inequality (\ref{stability}) does not ensure $\bar \ma >0, \bar \mc >0$ and $\bar \ma\bar \mc-\bar \mb^2>0$. If the matrix of kinetic coefficients is not positive definite, the effective dissipation $R_{int}$ that is ascribed to the interface within the phase field model may then be negative.
Two cases should thus be considered. When the conditions $\bar \ma >0, \bar \mc >0$ and $\bar \ma\bar \mc-\bar \mb^2>0$ are fulfilled, i.e. when the matrix of kinetic coefficients is positive definite, a direct comparison of the phase field simulations with time dependent calculations within the macroscopic approach using $\bar \ma$, $\bar \mb$ and $\bar \mc$ can be done. In the opposite case, one cannot make this comparison because the time dependent calculations within the macroscopic approach exhibit strong "unphysical" instabilities \cite{brener_temkin} (these instabilities do not exist in the underlying phase field model). Actually, the characteristic length scale $\lambda$ of the localized unstable mode is small, being of order $W$. It therefore does not fall into the range of applicability of the reduction procedure presented above, since the latter assumes that $W$ is much smaller than any macroscopic length scale of the diffusion field in the bulk.
Thus, this "unphysical" short length instability formally exists in the derived macroscopic description but does not appear in the underlying phase field model.
 It is easy to understand that indeed $\lambda \sim W$ using the simple situation of a steady diffusion flux across an immobile interface between two phases with the same diffusion coefficient $D$. In this case, the boundary conditions at the interface only involve the diffusional resistance $\bar \mc$. It may be shown using a linear stability analysis that, when $\bar \mc <0$, an unstable localized mode with a short length scale  $\lambda \sim D |\bar \mc|$ exists. In the phase field model, we have $\bar \mc <0$ when $D(\phi)$ exhibits a maximum in the interface region [see Eq. (\ref{barc})]. Since $D(\phi)>0$, the maximum magnitude of $|\bar \mc|$ in this case is of order $W/D$. We thus have $\lambda \sim W$. 
 
Although this strong short length scale instability prohibits direct numerical simulations within the macroscopic approach, one still may perform long wave length analytical calculations formally ignoring this  instability. Then these analytics may be compared with phase field simulations, and such a comparison will be presented in our study of step-bunching instability in the next section. 

The fact that the phase field model may be stable with the corresponding matrix of kinetic coefficients being non-positive definite shows that, in its simplest form, the macroscopic approach fails to fully describe the variety of situations allowed by the phase field model. This suggests that some interfaces in "exotic" materials may exhibit a non-positive definite effective Onsager matrix of macroscopic kinetic coefficients.

\subsection{Explicit results of the reduction procedure: constant diffusion coefficient and one-sided model} \label{examples_alloy}

Now we give the results of the reduction procedure presented in the previous paragraph for two models of binary alloys phase transformations: a model where the diffusion coefficient is constant and the one-sided model where the diffusion is neglected in the growing phase. For simplicity we assume that $\tau$ and $M$ are constant, i.e.
\be
\tau(\phi)=\tau \;\;\;\; \text{ and } \;\;\;\; M(\phi)=M.
\ee
In order to have explicit formulas for the equilibrium profiles $\phi_{eq}(x)$ and $C_{eq}(x)$, one should then choose the phase field potential $f(\phi)$ and the chemical free energy density $g(\phi,C)$. A usual choice for the phase field potential is a double-well potential of the form
\be \label{f}
f(\phi) = \phi^2 (1-\phi)^2,
\ee
for which we have
\be
\phi_{eq}(x) = \left\{1- \tanh \left[x/(\sqrt{2}W) \right] \right\}/2 \;. \nonumber
\ee
For the chemical free energy density, one may choose parabolic variations with the concentration $C$ \cite{plapp_folch}:
\be \label{g}
g(\phi,C) = \frac{1}{2}\; \Big[C-C_0^{eq} - q(\phi) (C_1^{eq} - C_0^{eq}) \Big]^2
\ee
with 
\be
q(\phi) = \phi^3 (10-15\phi+6\phi^2), \nonumber
\ee
leading to 
\be
C_{eq}(x) = (C_0^{eq}+C_1^{eq})/2 + v(x) (C_1^{eq}-C_0^{eq})/2 \nonumber
\ee
where $v(x) = -v(-x) = 2q[\phi_{eq}(x)] - 1$.

\subsubsection{Constant diffusion coefficient}

For a constant diffusion coefficient $D(\phi)=D$ we have, according to Eqs. (\ref{bara}), (\ref{barb}) and (\ref{barc}) where $D_1=D_0=D$:
\bea
\bar \ma &=& \frac{\alpha\tau}{W} - \frac{\beta W(C_1^{eq}-C_0^{eq})^2}{4D} - \alpha M (C_1^{eq}+C_0^{eq}), \nonumber \\
 \bar \mb &=& \alpha M, \nonumber \\
 \bar \mc &=& 0 ,\nonumber
\eea
 where
\bea
\alpha= W \int_{-\infty}^\infty dx [\phi'_{eq}(x)]^2 \approx 0.23570, \label{alpha} \\
 \beta = \int_{-\infty}^\infty \frac{dx}{W} [1-v^2(x)] \approx 1.40748 \nonumber \;.
\eea
The diffusional resistance of the interface $\bar \mc$ vanishes in this case. For $M=0$, one recovers the well-known results of the thin-interface limit \cite{karma_rappel} and its translation to the alloy problem with a concentration field \cite{elder}, where only $\bar \ma$ is non-vanishing. For $\tau < \beta (C_1^{eq}-C_0^{eq})^2 W^2/(4\alpha D)$, one then has a negative growth kinetic coefficient $\bar \ma <0$.
When the cross coupling $M \neq 0$ is introduced in the phase field equations, the diffusional resistance vanishes also but the cross coefficient $\bar \mb$ exists. One then have in all cases $\bar \ma\bar \mc-\bar \mb^2 <0$.   

Note that the expressions for $\bar \ma, \bar \mb$ and $\bar \mc$ given above are equivalent to Eqs. (22), (23) and (24) in Ref. \cite{brener_boussinot} with $\delta=0$. This illustrates the difference of reduction procedure performed here in comparison with Ref. \cite{brener_boussinot} as described in the last paragraph. More precisely, the subtraction of the last two terms in Eq. (\ref{rint_pf1}), that is not performed in Ref. \cite{brener_boussinot}, leads to $\delta$-independent kinetic coefficients here.

\subsubsection{One-sided model for binary alloy solidification}

In the one-sided model for solidification of binary alloys, the phase 1 (the solid) corresponding to $\phi=1$ grows at the expense of phase 0 (the liquid) corresponding to $\phi=0$. The diffusion coefficient in phase 1 is much smaller than in phase 0, i.e. $D_1 \ll D_0$. In this case, the diffusional flux in phase 1 is small and can be neglected. According to Eq. (\ref{flux1}), we thus have, close to equilibrium: 
\be
J_B = V C_1^{eq}. \nonumber
\ee
The differences of chemical potentials then read
\bea
\delta \mu_A = (\bar \ma + \bar \mb C_1^{eq}) V,  \nonumber\\
\delta \mu = (\bar \mb + \bar \mc C_1^{eq}) V . \nonumber
\eea
According to Eqs. (\ref{bara}), (\ref{barb}) and (\ref{barc}), the combinations $\bar \ma + \bar \mb C_1^{eq}$ and $\bar \mb + \bar \mc C_1^{eq}$ are independent of $D_1$, and therefore the differences of chemical potentials only depend on the diffusion coefficient in the liquid $D_0$.
We write the phase field dependence of the diffusion coefficient as 
\be \label{diff}
D(\phi)=D_0 (1-\phi),
\ee
 such that $D(\phi=1)=0$ in phase 1 and $D(\phi=0)=D_0$ in phase 0. 
According to Eqs. (\ref{barb}) and (\ref{barc}) we have 
\be \label{b+cc1}
\bar \mb + \bar \mc C_1^{eq} = \alpha M - \frac{(C_0^{eq}-C_1^{eq})W\rho}{2D_0} ,
\ee
where $\alpha$ is given in Eq. (\ref{alpha}) and 
\be
\rho = \int_{-\infty}^\infty \frac{dx}{W} \; \frac{\phi_{eq}(x)-v(x)}{1-\phi_{eq}(x)}  \approx 2.12132 \;. \nonumber
\ee
%$\bar \mb + \bar \mc C_1^{eq}$ is convergent because $1-u(x)$ converges to 0 faster than $1-\phi_{eq}(x)$ when $\phi_{eq} \to 1$. 
According to Eqs. (\ref{bara}) and (\ref{barb}) we have
\bea
\bar \ma + \bar \mb C_1^{eq} &=& \frac{\alpha \tau}{W} - \frac{(C_0^{eq}-C_1^{eq})^2 W\zeta}{4D_0} \nonumber\\
&&-  C_0^{eq} \Big[ \bar \mb + \bar \mc C_1^{eq} \Big] ,
 \label{a+bc1}
 \eea
where
\be
\zeta = \int_{-\infty}^\infty \frac{dx}{W} \; \frac{1-v^2(x)}{1-\phi_{eq}(x)} \approx 3.42778 \;. \nonumber
\ee

%{\bf Cancellation of $\delta \mu$ and $\delta \mu_A$.} 
{\it{Anti-trapping model.}} In the phase field modeling of solidification, it is often assumed that there is no jump of diffusion chemical potential at the interface, i.e. $\delta \mu=0$.
Here this is provided by:
\bea
M=M^* = \frac{(C_0^{eq}-C_1^{eq})W\rho}{2\alpha D_0}  \nonumber,
\eea
so that the r-h-s of Eq. (\ref{b+cc1}) vanishes.
The anti-trapping current ${\bf J_{at}}$ \cite{antitrapping, fang},
such that [see Eq. (\ref{diff_flux})] 
\be
{\bf J} =  -D(\phi) \boldnabla \frac{\delta G}{\delta C} + {\bf J_{at}}  \;,\nonumber
\ee
then reads
\bea
{\bf J_{at}} &=& -D(\phi) M^*W \dot \phi \boldnabla \phi \nonumber \\ 
&=& -\frac{\rho (C_0^{eq}-C_1^{eq})}{2\alpha} \; (1-\phi) W^2 \dot \phi \boldnabla \phi \nonumber .
\eea
%In Ref. \cite{fang}, a "mesoscopic solute drag" effect is defined such that the vector, having the dimension of a length, ${\bf m} = D(\phi)MW \boldnabla \phi$ verifies
%\be
%\dot \phi \left[1-\frac{{\bf m}^2}{D(\phi) \tau}) \right] = -\frac{1}{\tau} \left[ \frac{\delta G}{\delta \phi} - {\bf m} \cdot \boldnabla \frac{\delta G}{\delta C} \right].
%\ee
%Here, with the choice $M=M^*$, we have
%\be
%{\bf m}^* = \frac{\rho}{2\alpha} (C_0^{eq}-C_1^{eq}) (1-\phi)W^2 \boldnabla \phi.
%\ee
In this frame, $\delta \mu_A$ is obtained through Eq. (\ref{a+bc1}) with $M=M^*$, i.e. $\bar \mb + \bar \mc C_1^{eq}=0$:
\be
\delta \mu_A = \left[\frac{\alpha \tau}{W} - \frac{(C_0^{eq}-C_1^{eq})^2 W \zeta}{4D_0}\right] V  \;. \nonumber
\ee
%In this model for $f(\phi)$ where $[W\phi'_{eq}(x)]^2 = 2 \phi_{eq}^2 (1-\phi_{eq})^2$, we have
%\bea
%\text {max} \left\{D(\phi) [W\phi'(x)]^2 \right\} &=& 2D_0\text {max}  \left\{\phi^2 (1-\phi)^3 \right\} \nonumber \\
%&=& \frac{216}{3125} D_0 \nonumber
%\eea
%for $\phi=2/5$.
%The inequality (\ref{stability}) implies
%\be
%D_0 < \frac{3125}{216 } \; \frac{\tau}{M^{*2}}, \nonumber
%\ee
%which means 
%\be
%(\bar \ma + \bar \mb C_1^{eq})^* > \frac{\alpha \tau}{W} \left[ 1- \frac{3125}{216} \frac{\alpha \zeta}{\rho^2}\right]=- \eta \tau/W \nonumber
%\ee
%where $\eta \approx 0.37653$. One may then choose 
In addition to $\delta \mu=0$, one may require also $\delta \mu_A=0$ in order to fully eliminate kinetic effects at the solidification front. This is provided by the choice
\be
\tau = \tau^* = \frac{\zeta(C_0^{eq}-C_1^{eq})^2 W^2}{4\alpha D_0} \nonumber
\ee
that lies in the range of stability of the phase field model since the inequality (\ref{stability}) with $M=M^*$ and $\tau=\tau^*$ holds.

 Therefore the equations of motion (\ref{phase_field}) and (\ref{diff_flux}) with a diffusion coefficient given by Eq. (\ref{diff}) and where the free energy functional in Eq. (\ref{free_energy}) is specified by Eqs. (\ref{f}) and (\ref{g}) gives $\delta \mu=0$ at the solidification front when $M=M^*$ and in addition gives $\delta \mu_A=0$ when $\tau=\tau^*$.

%Replacing the solid equilibrium concentration $C_1^{eq}=C_S$ and the liquid equilibrium concentration $C_0^{eq}=C_L$, 
%one may then summarize the phase field model for solidification of a binary alloy with equilibrium boundary conditions as follows
%\bea
%[1-\rho^2\alpha(1-\phi)(W\boldnabla \phi)^2/\zeta] \dot \phi &=& \nonumber \\
%-\frac{4\alpha D_0/W^2}{\zeta (C_L-C_S)^2 } \frac{\delta G}{\delta \phi}  
%+ \frac{2\rho \alpha D_0(1-\phi) }{\zeta(C_L-C_S)} \boldnabla \phi \cdot \boldnabla \frac{\delta G}{\delta C}
%\eea
%and
%\bea
%\dot C &=& D_0 \boldnabla \cdot \left[ \frac{(1-\phi)\boldnabla (\delta G/\delta C)}{1-\rho^2\alpha(1-\phi)(W\boldnabla \phi)^2/\zeta}   \right] \\
%&& + \frac{2\rho \alpha D_0 }{\zeta(C_L-C_S)} \boldnabla \cdot \left[ \frac{(1-\phi) (\delta G/\delta \phi) \boldnabla \phi}{1-\rho^2\alpha(1-\phi)(W\boldnabla \phi)^2/\zeta}  \right] \nonumber
%\eea
% where
%\bea
%G[C,\phi]=\int dV \Big\{H \Big[W^2 (\boldnabla\phi)^2/2+\phi^2(1-\phi)^2 \Big] \\
% + \Big[C-C_L + \phi^3(10-15\phi+6\phi^2)(C_L-C_S) \Big]^2 \Big/2\Big\} \;.
%\eea
%\\

%In these two examples of phase field models for binary alloy phase transitions, we gave the link between the parameters of the phase field model and the kinetic boundary conditions in the  macroscopic approach. The originality mostly comes from the simplicity of the derivation (in comparison with the asymptotic matching procedure in the thin interface limit) and the fact that we have given in Eqs. (\ref{bara}), (\ref{barb}) and (\ref{barc}) general formulas for the kinetic coefficients. 

\section{Steps dynamics in molecular-beam-epitaxy (MBE)} \label{mbe}

We now present our study of steps dynamics in MBE. We first give  a brief overview of the macroscopic description of steps dynamics, and especially the kinetic boundary conditions at the steps (for more details, we refer to Ref. \cite{misbah}). Second we present the phase field model and the corresponding kinetic boundary conditions derived using the reduction procedure presented in general terms in the previous section. We finally perform simulations of the presented phase field model for the step-bunching instability and compare the numerical results to analytical predictions within the macroscopic approach.

\subsection{Macroscopic description of steps dynamics}

In MBE, one usually assumes that the adatom concentration $c$ on a terrace obeys a two-dimensional diffusion equation:
\be\label{diff_eq_mbe}
\dot c = D \boldnabla^2 c + F - c/\tau_v \;,
\ee
where $F$ is the flux of adatom from the beam and $\tau_v$ is a characteristic time for desorption of adatoms from the terrace back to the vapor. The concentration $c$ represents the surface density of diffusing adatoms on a terrace and $c=1$ corresponds to the surface density of the crystal.
At equilibrium, the adatom concentration on the terraces is constant and equal to $c_{eq}$. When $F\tau_v - c_{eq} \neq 0$, the system is driven out of equilibrium and gradients of concentration appear on the terraces. 
At a step, the mass conservation is written:
\be\label{mass_cons}
V=D(\boldnabla c|_+ - \boldnabla c|_-)\cdot {\bf n},
\ee
where $\bf n$ is the unit vector normal to the step,  $V$ is the normal velocity of the step and $\boldnabla c|_+$ ($\boldnabla c|_-$) is the concentration gradient at the step on the side of the lower (upper) terrace. 
%This equation is correct in two cases, i.e. if kinetic effects are small (in the local equilibrium approximation) or if $F\tau_v - c_{eq} \ll 1$ (in the static approximation where $\dot c$ is neglected). 
Two macroscopic length scales are present in this problem, i.e. $\sqrt{D\tau_v}$ and $D/V$. When $F\tau_v - c_{eq} \ll 1$, one may use the quasi-static approximation, $\dot c=0$. One then has $D/V \gg \sqrt{D\tau_v}$. 

\begin{figure}[htbp]
\includegraphics[width=\linewidth]{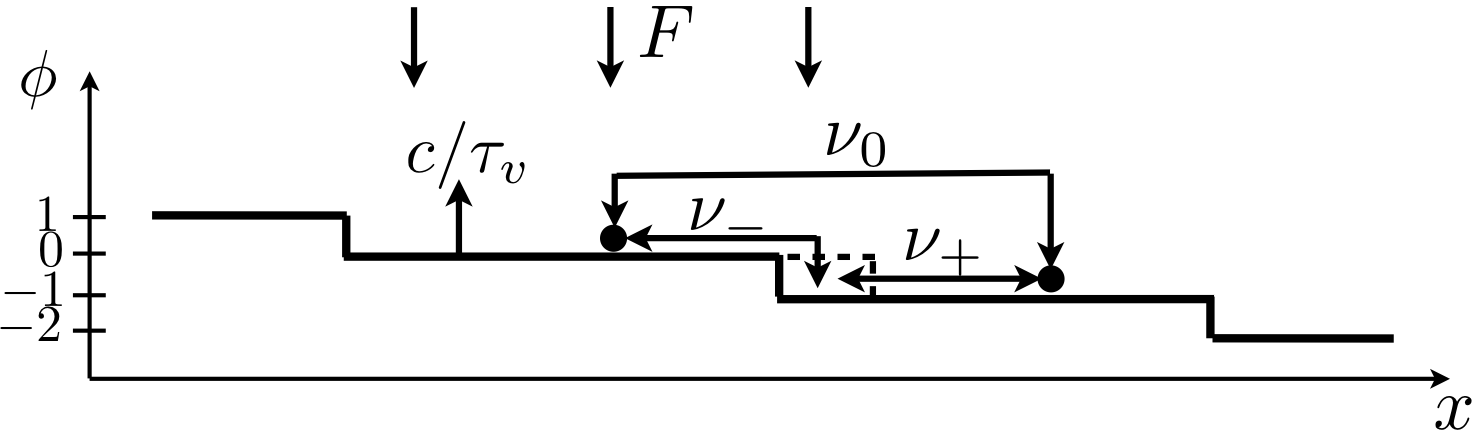}
\caption{\label{schema_mbe} Schematic representation of the physical mechanisms at play for steps dynamics in molecular-beam-epitaxy (MBE). The flux $F$ from the beam produces an out-of-equilibrium concentration $c$ of adatoms on the terraces. These adatoms diffuse on the terraces, and may desorb back to the vapor with a characteristic time $\tau_v$. The dashed line represents the advancement of the step due to the attachment of adatoms from the lower terrace (described by the kinetic coefficient $\nu_+$) and the attachment of adatoms from the upper terrace (described by the kinetic coefficient $\nu_-$). The third kinetic coefficient $\nu_0$ describes atomic exchanges between terraces, without attachment to the step. Integer values of the phase field $\phi$ quantify the height of the crystal in atomic units.}
\end{figure}

In absence of kinetic effects, the adatom concentration at the step $c_+$ ($c_-$)  on the lower (upper) terrace  is equal to $c_{eq}$. When attachment kinetics are relevant, $c_+$ and $c_-$ are related to the diffusional fluxes by the kinetic boundary conditions:
\bea \label{bc_mbe1}
D \boldnabla c|_+ \cdot {\bf n} = \nu_+ X_+ + \nu_0 (X_+ - X_-), \\
- D \boldnabla c|_- \cdot {\bf n} = \nu_- X_- + \nu_0 (X_- - X_+),  \label{bc_mbe2}
\eea
where $X_\pm = c_\pm - c_{eq}$. The kinetic coefficient $\nu_+$ ($\nu_-$) describes the attachment of adatoms from the lower (upper) terrace to the step, while $\nu_0$, called step transparency \cite{transparency}, describes atomic exchanges between terraces without attachment to the step (see Fig. \ref{schema_mbe}). 
%For a detailed description of these boundary conditions we refer to the review article \cite{misbah}.

The possibility of having $\nu_+ \neq \nu_-$ is called the Ehrlich-Schwoebel (ES) effect \cite{ehrlich,schwoebel}. It accounts for the difference of energy barrier for adatoms to attach to the step whether coming from the lower or from the upper terrace. The ES effect is responsible for instabilities such as step-bunching \cite{schwoebel} or meandering \cite{bales}.

{\it Analogy with the binary alloy problem.} By subtracting Eq. (\ref{flux2}) from Eq. (\ref{flux1}) with $D_1=D_0=D$ and $C_1=C_0+1$, one recovers Eq. (\ref{mass_cons}). Hence the problem of steps dynamics on a vicinal surface is mathematically equivalent to a problem of phase transition in a "binary A-B alloy" with an unbounded concentration $C$ of atoms B. If one considers a distribution of steps at the surface of a crystal and $n$ denotes the height (in units of the atomic distance) of the crystal perpendicularly to its surface, each $n$ represents a different phase of the "binary alloy" with an equilibrium concentration $C_n^{eq}$ such that $C_n^{eq} = C_{n-1}^{eq}+1$. The thermodynamic equilibrium then consists of a mixture of these phases, and out of equilibrium the concentration $C$  in the $n$-th phase differs from $C_n^{eq}$. The link between $C$ and the adatom concentration $c$ on the $n$-th terrace is provided by 
\be
c-c_{eq} = C-C_n^{eq} \;, \nonumber
\ee
with $C_n^{eq} = c_{eq}+n$.

\subsection{Phase field model} \label{model_mbe}

In this section we present the phase field model for steps dynamics in MBE. We give explicit choices for the phase field potential, the chemical free energy density and the diffusion coefficient. We then use the analogy with the binary alloy problem presented just above to write down the phase field equations of motion.
 
We use the phase field $\phi$ to quantify the height of the crystal, with $\phi=n$ on the terraces (see Fig. \ref{schema_mbe}). We use a periodic potential 
\be
f(\phi) = [1-\cos(2\pi \phi)]/(2 \pi) \nonumber
\ee
with $f(\phi+1) = f(\phi)$. The equilibrium phase field profile $\phi_{eq}(x)$ has no explicit expression but obeys 
\be \label{eq_pf}
[W\phi'_{eq}(x)]^2= 2f[\phi_{eq}(x)] .
\ee
We use a dimensionless chemical free energy density of parabolic form
\be
g(C,\phi) = \big[C-c_{eq}-p(\phi) \big]^2/2 ,\nonumber
\ee
with 
\be
p(\phi) = \phi - \sin(2 \pi \phi)/(2 \pi) \nonumber
\ee
 that has the property $p(\phi=n) = n$ and $p'(\phi=n)=p''(\phi=n)=0$. The equilibrium profile of $C$ is 
\be \label{eq_c}
C_{eq}(x) = c_{eq} + p[\phi_{eq}(x)] .
\ee
One then defines an adatom concentration 
\be
c = C - p(\phi) \; \nonumber
\ee
that is continuous across a step and takes a constant value equal to $c_{eq}$ at equilibrium.

For simplicity we assume that the time scale $\tau(\phi)$ and the inverse velocity scale $M(\phi)$ are constant:
\be
\tau(\phi)=\tau \;\;\;\; \text{ and } \;\;\;\; M(\phi)=M.
\ee
 Moreover, all terraces are thermodynamically equivalent and therefore the diffusion coefficient on the different terraces is the same:
\be
D(\phi=n) = D.
\ee
 However we introduce a variation of the diffusion coefficient within the interface characterized by the dimensionless parameter $g_{ {}_D}$:
\be \label{diff_coeff_mbe}
D(\phi) = \frac{D}{1+g_{ {}_D} (W\boldnabla \phi)^2} \;.
\ee
We will see later that the introduction of a finite $g_{{}_D}$ is crucial for the development of the step-bunching instability due to the ES effect within our model. The constraint $D(\phi)>0$ implies $g_{ {}_D}>$ -1/max$[(W\boldnabla \phi)^2]$. The maximum value of $(W\boldnabla \phi)^2$ at equilibrium is max$[(W\phi'_{eq})^2] = 2/\pi$. Therefore, close to equilibrium, the conditions of stability of our phase field model read
\bea 
\tau>0 \; ; \; D>0 \; ; \; g_{ {}_D} > - \pi/2 \; ; \; M^2 < (\pi /2 + g_{ {}_D}) \tau/D \;. \nonumber\\ \label{stability_mbe}
\eea

Finally, for the phase field model of step dynamics in MBE detailed above,  Eqs. (\ref{phase_field}) and (\ref{diff_flux}) read
\bea \label{phase_field_mbe}
 -\frac{\delta G}{\delta \phi} &=& \tau \dot \phi +   ( MW \boldnabla \phi) \cdot {\bf J} ,\\
-\boldnabla \frac{\delta G}{\delta C} &=& ( MW \boldnabla \phi) \; \dot \phi  + \frac{1+g_{ {}_D}(W\boldnabla \phi)^2 }{D} \;{\bf J} \label{diff_flux_mbe} , 
 \eea
with 
\bea
G[C,\phi]=\int dV \Big\{H \left[\frac{(W\boldnabla\phi)^2}{2}+ \frac{1-\cos(2\pi \phi)}{2 \pi}\right] \nonumber \\ 
+\big[C-c_{eq}-p(\phi)\big]^2/2\Big\} \;, \label{free_energy_mbe}
\eea
and they are subjected to the inequalities (\ref{stability_mbe}).
The continuity equation that takes into account the flux from the beam and the desorption effect reads
\be
\dot C = - \boldnabla \cdot {\bf J} + F - \frac{C-p(\phi)}{\tau_v} .\label{continuity_mbe}
\ee
In the bulk where $\dot \phi=0$, $\boldnabla \phi=0$ and $\delta G/\delta C = \partial g/\partial C=C-c_{eq}-p(\phi)=c-c_{eq}$, we recover the diffusion equation (\ref{diff_eq_mbe}).

\subsection{Relation between phase field parameters and kinetic boundary conditions (reduction procedure)}

The couples of conjugated fluxes and driving forces that are commonly used to describe kinetic boundary conditions at a step in MBE are different than those that are commonly used to describe phase transformations in binary alloys. In MBE, the diffusional flux on the lower side (upper side) of the step $\boldnabla c|_+ \cdot {\bf n}$ ($-\boldnabla c|_- \cdot {\bf n}$) is conjugated to the deviation from equilibrium concentration on this side $X_+ = c_+ - c_{eq}$ ($X_- = c_- - c_{eq}$). These two couples of conjugated quantities are linearly related through the Onsager matrix \cite{pierre_louis}:
\bea \label{kinetic+}
D \boldnabla c|_+\cdot {\bf n} &=& L_+ X_+ + L_0 X_- \;,\nonumber \\
-D \boldnabla c|_-\cdot {\bf n} &=& L_0 X_+ + L_- X_- \;. \label{kinetic-} \nonumber
\eea
According to Eqs. (\ref{bc_mbe1}) and (\ref{bc_mbe2}), one thus has
\bea
L_\pm = \nu_\pm + \nu_0, \nonumber \\ 
L_0 = -\nu_0 . \nonumber
\eea

We now have to express the correspondence between the difference of chemical potentials $\delta \mu_A$ and $\delta \mu$ presented in the previous section and the driving forces $X_+$ and $X_-$. We consider a step that connects phase 0 $(\phi=0, C_0^{eq}=c_{eq})$ to phase 1 $(\phi=1, C_1^{eq}=c_{eq}+1)$. Then $X_+ = C_0-C_0^{eq}$ and $X_- = C_1-C_1^{eq}$.
Since the terraces are thermodynamically equivalent, the dimensionless free energy density $g_1(C)$ of phase 1 and $g_0(C)$ of phase 0 are such that $g_1(C) = g_0(C-1)$. Therefore their second derivative that enters the driving forces $\delta \mu_A$ and $\delta \mu$ close to equilibrium [Eqs. (\ref{delta_mu_exp})] are equal, i.e. $g''_1(C_1^{eq})=g''_0(C_0^{eq})$, and are set to 1 for convenience. The driving forces are thus
\bea
\delta \mu_A &=& \bar \ma V + \bar \mb J_B =-c_{eq} X_+ + (c_{eq}+1) X_-  \nonumber \\
\delta \mu &=& \bar \mb V + \bar \mc J_B=X_+ - X_- \nonumber
\eea
Using the macroscopic boundary conditions close to equilibrium [see Eqs. (\ref{flux1}) and (\ref{flux2})]
\bea
D \boldnabla c|_+ \cdot {\bf n} &=& - V c_{eq} + J_B  \;, \nonumber\\
-D \boldnabla c|_- \cdot {\bf n} &=&  V (c_{eq}+1) - J_B \;, \nonumber
\eea
we obtain
\bea
L_+ (\bar \ma \bar \mc - \bar \mb^2) = \bar \ma + c_{eq} [2 \bar \mb + \bar \mc c_{eq}], \label{Lp} \nonumber \\
-L_0 (\bar \ma \bar \mc - \bar \mb^2) = \bar \ma + \bar \mb + c_{eq}[2\bar \mb + \bar \mc (c_{eq}+1) ]  , \label{L0} \nonumber\\
L_- (\bar \ma \bar \mc - \bar \mb^2) =   \bar \ma + (c_{eq}+1)[2\bar \mb + \bar \mc (c_{eq}+1) ]   . \label{Lm}
\eea
The positive definiteness of this Onsager matrix requires $L_+>0$, $L_->0$ and $L_+ L_- > L_0^2$. One may easily check that $(L_+ L_- -L_0^2 )(\bar \ma \bar \mc - \bar \mb^2)= 1$. 

At this point we have linked the representations ($\delta \mu_A,\delta \mu;V,J_B$) and ($X_+,X_-;D \boldnabla c|_+\cdot {\bf n},-D \boldnabla c|_-\cdot {\bf n}$). This formal link is actually allowed by the fact that the contributions from the flux $F$ and from the desorption term in the integration across the step (of width $W$) of the continuity equation (\ref{continuity_mbe}) are negligible with respect to the other contributions. It is clear for example in the case of an isolated straight step. Its steady-state velocity $V$, which is the order of magnitude of the integral over $W$ of $\dot C$ or $\boldnabla \cdot \bf J$, is of order $V \sim (F-c_{eq}/\tau_v)\sqrt{D\tau_v} \gg (F-c_{eq}/\tau_v)W$.

%The kinetic coefficients that are usually used to discuss step dynamics are 
%\bea
%\nu_\pm &=& L_\pm + L_0 \label{nupm}\\
%\nu_0 &=& -L_0 \label{nuo}
%\eea
%where $\nu_+$ ($\nu_-$) describes the attachment to the step of adatoms of the lower (upper) terrace and $\nu_0$ is called the step transparency coefficient describing jumps of adatoms from one terrace to the other. Then the normal velocity of the steps is $V=\nu_+ X_+ + \nu_- X_-$.
%The Ehrlich-Schwoebel (ES) effect consists in having $\nu_+-\nu_- \neq 0$ describing a different energetic barrier for an adatom to attach to the step whether coming from the upper or the lower terrace.

One should note that in the above derivation, if one considers an interface between phases $n$ and $n+1$ of the "binary alloy", one just replaces $c_{eq}$ in the above equations by $c_{eq}+n$. However, we will see below, when we give the kinetic coefficients $\nu_+, \nu_-,\nu_0$ in terms of the parameters of the phase field model, that the dependency on $c_{eq}$ and $n$ is removed. This is necessary since the choice of $n$ is arbitrary.

When one inserts the characteristics of the equilibrium distributions, that are obeying Eqs. (\ref{eq_pf}) and (\ref{eq_c}), and the diffusion coefficient [Eq. (\ref{diff_coeff_mbe})] into Eqs. (\ref{bara}), (\ref{barb}) and (\ref{barc}), we obtain
\bea
\bar \ma = \frac{\alpha \tau}{W} - \frac{\beta W (C_1^{eq} - C_0^{eq})^2}{4D} -\alpha M (C_1^{eq} + C_0^{eq})\nonumber \\
 + \left[\frac{(C_1^{eq} + C_0^{eq})^2}{4} + \frac{(C_1^{eq} - C_0^{eq})^2 }{4} \; \frac{\chi}{\alpha} \right]\; \bar \mc, \nonumber\\
 \bar \mb = \alpha M - \frac{C_1^{eq} + C_0^{eq}}{2} \; \bar \mc , \nonumber\\
 \bar \mc = \frac{\alpha Wg_{{}_D}}{D}, \label{barc_mbe}
\eea
where the numerical factors are 
\bea
 \alpha &=& W \int_{-\infty}^ \infty dx [\phi_{eq}'(x)]^2 =  \int_0^1 d\phi \sqrt{\frac{1-\cos(2\pi \phi)}{\pi}} \nonumber \\
 &=& (2/\pi)^{3/2} , \nonumber\\
\beta &=& \int_{-\infty}^ \infty \frac{dx}{W} [1-u^2(x)]  = 4 \sqrt{\pi} \int_0^1 d\phi \; \frac{p(\phi)[1-p(\phi)]}{\sqrt{1-\cos(2\pi \phi)}} \nonumber\\
& \approx& 0.60595 , \nonumber\\
\chi &=&W \int_{-\infty}^ \infty dx \; [\phi_{eq}'(x)]^2 u^2(x) \nonumber\\
&=& \int_0^1 d\phi  \; \Big\{1-4p(\phi) \big[1-p(\phi) \big] \Big\} \sqrt{\frac{1-\cos(2\pi \phi)}{\pi}}  \nonumber \\
&\approx& 0.21516 ,\nonumber
\eea
with $u(x)=-u(-x)=1-2p[\phi_{eq}(x)]$. 
With $C_1^{eq}=c_{eq}+1$ and $C_0^{eq}=c_{eq}$ and the relations in Eqs. (\ref{Lm}), the kinetic coefficients for the steps dynamics in MBE are
\bea
 \nu_\pm &=& \left[ \mp \alpha M + \frac{\alpha Wg_{ {}_D}}{2D}  \right] \Delta^{-1} \label{nupm}\\
\nu_0 &=& \left[ \frac{\alpha \tau}{W} - \frac{\beta W}{4D} + \frac{(\chi-\alpha)Wg_{ {}_D}}{4D}  \right] \Delta^{-1}  \label{nuo}
\eea
where
\bea \label{delta}
\Delta &=& \bar \ma \bar \mc - \bar \mb^2 = \big[\nu_+ \nu_- + \nu_0 (\nu_+ + \nu_-) \big]^{-1} \nonumber\\
&=& -\alpha^2 M^2 + \frac{\alpha Wg_{ {}_D}}{D} \left[ \frac{\alpha \tau}{W} - \frac{\beta W}{4D} + \frac{\chi W g_{ {}_D}}{4D} \right]
\eea
We note that the expressions for $\nu_\pm$ and $\nu_0$ are indeed independent of $c_{eq}$ and therefore of $n$. We see moreover that the ES effect is due to the cross terms in the equations of motion (\ref{phase_field_mbe}) and (\ref{diff_flux_mbe}) that are parametrized by $M$, i.e. $(\nu_+-\nu_-)\Delta = -2\alpha M$.

\section{Simulation of the step-bunching instability due to Ehrlich-Schwoebel effect}

\begin{figure}[htbp]
\includegraphics[width=\linewidth]{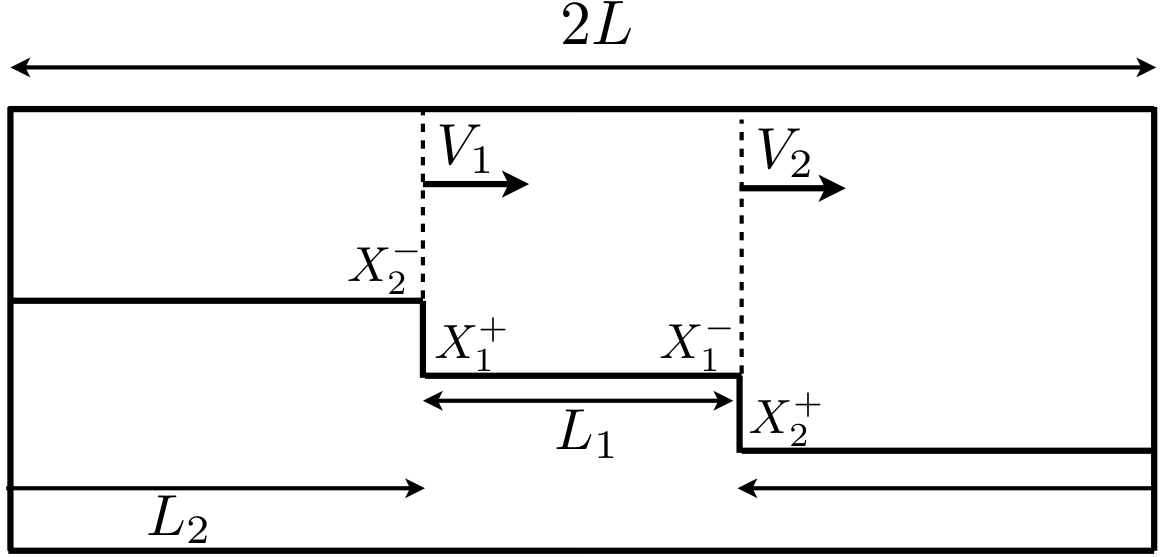}
\caption{\label{schema_pairing} Pair of steps that is simulated numerically. The mean distance $L$ between steps  is time independent and $L_1$ and $L_2$ are time dependent due to the time dependent velocities $V_1$ and $V_2$ of the two steps. $X_1^\pm$ and $X_2^\pm$ are the deviations from equilibrium concentrations (see Appendix \ref{macro_steps}).}
\end{figure}

We have investigated numerically the step-bunching instability due to the ES effect in MBE \cite{schwoebel}. We consider a vicinal surface on which a train of parallel equidistant steps (step-flow regime) may be unstable and we investigate the mode for which parallel steps are forming pairs. 
For the simulations, two steps are present in a one-dimensional simulation box described by the coordinate $x$. The length of the simulation box is $2 L$ where $L$ is the average distance between the steps. The boundary conditions at the borders of the simulation box are such that $\phi(x=0) = \phi(x=2L)+2$ and $C(x=0) = C(x=2L)+2$. This corresponds to periodic boundary conditions for the adatom concentration $c(x=0)=c(x=2L)$.
We define the time dependent quantity $\epsilon(t)$ such that the distance $L_1(t)=[1-\epsilon(t)]L$ between two steps decreases ($\dot \epsilon>0$) when $V_1(t)-V_2(t)=L\dot \epsilon >0$, where $V_1(t)$ and $V_2(t)$ are the time dependent step velocities (see Fig. \ref{schema_pairing}).
 Initially, $\epsilon$ is set to a small positive value, and we measure $\epsilon(t)$ in the course of the simulation. The rate $\lambda = \dot \epsilon(t)/\epsilon(t)$ describes a relaxation to the step flow regime for $\lambda <0$ and describes an instability for $\lambda >0$. 

In the step-flow regime, the vicinal surface is stable with $\epsilon=0$ and $V_1(t)=V_2(t) = V$. 
We refer to the Appendix \ref{macro_steps} for a derivation of $V$.
%It is shown within the macroscopic approach in Appendix \ref{macro_steps} that the steady-state velocity is, assuming a small driving force $F\tau_v - c_{eq} \ll 1$, 
%\bea
%V &=& V_{eq} \; \frac{1+ J_v \Delta \frac{\nu_+ + \nu_-}{2} \frac{\cosh(\sigma) + 1}{\sinh(\sigma)} }{1+ 2 J_v \Delta \nu_0 \frac{\cosh(\sigma) - 1}{\sinh(\sigma)} + J_v \Delta \frac{\nu_+ + \nu_-}{\tanh(\sigma)} + J_v^2 \Delta } \nonumber
%\eea
%where
%\bea
%J_v &=& \sqrt{\frac{D}{\tau_v}} , \nonumber\\
%\sigma &=& \frac{L J_v}{D},
%\eea
%and 
%\bea
%V_{eq} = 2 J_v  \frac{\cosh(\sigma) - 1}{\sinh(\sigma)}  (F\tau_v - c_{eq}) \nonumber
%\eea
%is the steady-state velocity in the case of equilibrium boundary conditions, i.e. without kinetic effects.
When $\epsilon \neq 0$, one has $V_1 \neq V_2$ and the system whether relaxes to the step-flow regime with $\dot \epsilon/\epsilon <0$ or exhibits the pairing instability with $\dot \epsilon/\epsilon >0$. If one assumes, in addition to $F\tau_v - c_{eq} \ll 1$, that kinetic effects are small, i.e. when the $\nu's$ are much larger than the velocity scales $\sqrt{D/\tau_v}$ and $D/L$, one may obtain analytically within the macroscopic approach (see Appendix \ref{macro_steps})
\bea 
\lambda &=& -\frac{2 D}{\tau_v^2 \sinh^2\sigma} (\nu_+^2 - \nu_-^2) \Delta^2 (F\tau_v - c_{eq}) \nonumber \\
&& + \frac{4 (\cosh \sigma -1) (\sinh \sigma - \sigma \cosh \sigma)}{\tau_v \sinh^3 \sigma} (F\tau_v - c_{eq})^2, \nonumber\\ \label{lambda}
\eea
where $\sigma = L/\sqrt{D\tau_v}$.
In terms of the phase field parameters, this gives, using Eqs. (\ref{nupm}) and (\ref{nuo}),
\bea \label{lambda_pf}
\lambda(M,g_{ {}_D})  &=& \frac{4 \alpha^2MW g_{ {}_D}}{\tau_v^2 \sinh^2\sigma}  (F\tau_v - c_{eq}) \nonumber \\
&& + \frac{4 (\cosh \sigma -1) (\sinh \sigma - \sigma \cosh \sigma)}{\tau_v \sinh^3 \sigma} (F\tau_v - c_{eq})^2 .\nonumber\\
\eea
In the following, we compare the rate $\lambda$ resulting from the phase field simulations to the one obtained within the macroscopic approach. Unfortunately, the solution for $\lambda$ converges very slowly to Eq. (\ref{lambda_pf}) when $\tau_v \to \infty$ and $L \to \infty$ and being in this limit implies a prohibitive computational cost (especially due to the length $2L$ of the simulation box). We will therefore compare the rate resulting from phase field simulations with a semi-analytical one computed numerically using the procedure described in the Appendix \ref{macro_steps}. However, it is very insightful to analyze the structure of Eq. (\ref{lambda_pf}) because it provides, as we will see later on, the qualitative behavior of $\lambda$ in the regime that was investigated with phase field simulations.
%In the following, we compare the rate $\lambda$ resulting from the phase field simulations to the one computed numerically using the procedure described in the Appendix \ref{macro_steps}. One may however obtain an analytical formula for $\lambda$ when, in addition to the assumption $F\tau_v - c_{eq} \ll 1$, one assumes that kinetic effects are small, i.e. when the $\nu's$ are much larger than the velocity scales $J_v$ and $D/L$. 
%
%It can be shown using the procedure given in the Appendix \ref{macro_steps} that, in the case where $\sqrt{D/\tau_v} \ll \nu_\pm$ and $\sqrt{D/\tau_v} \ll \nu_0$, one has
%\bea \label{lambda}
%\lambda &=& -\frac{2 D}{\tau_v^2 \sinh^2\sigma} (\nu_+^2 - \nu_-^2) \Delta^2 (F\tau_v - c_{eq}) \nonumber \\
%&& + \frac{4 (\cosh \sigma -1) (\sinh \sigma - \sigma \cosh \sigma)}{\tau_v \sinh^3 \sigma} (F\tau_v - c_{eq})^2, \nonumber\\
%\eea
%where $\sigma=L/\sqrt{D\tau_v}$.   
%{\color{blue} In the following we will compare simulations with analytics when we vary $M$ and $g_{{}_D}$ in the phase field model. Unfortunately, CPU time restrictions prohibit us to reach the regime $\sqrt{D/\tau_v} \ll \nu_\pm$ and $\sqrt{D/\tau_v} \ll \nu_0$. However, it is interesting to examine the structure of Eq. (\ref{lambda_pf}) that gives the qualitative behavior of the theory.}
 The first term on the r-h-s of Eq. (\ref{lambda_pf}) describes the kinetic effects. It is obtained in the static approximation where $\dot c$ is neglected in Eq. (\ref{diff_eq_mbe}). It is proportional to the driving force $F\tau_v - c_{eq}$ and to $Mg_{ {}_D}$. Therefore the ES effect $(\nu_+ - \nu_-)\Delta$, proportional to $M$, is not the sole ingredient for the instability to occur. A diffusional  resistance [coefficient $\bar \mc$ proportional to $g_{ {}_D}$, see Eq. (\ref{barc_mbe})] of the step is also required.
% This therefore shows that the kinetic cross coupling (parametrized by $M$) in the present phase field model is required to obtained this instability. Moreover it shows that a "Kapitza resistance" (coefficient $\bar \mc$ parametrized by $g_{ {}_D}$) of the interface is also required for this instability to occur in this phase field model. 
  The second term on the r-h-s of Eq. (\ref{lambda_pf}) does not contain kinetic coefficients and is present in the case of equilibrium boundary conditions. It accounts for the convective correction to the concentration field on the terraces due to $\dot c$. We refer to the Appendix \ref{macro_steps} for more details. It is proportional to $(F\tau_v - c_{eq})^2$, is negative and promotes the stability of the step flow regime. 
For the instability to occur, the magnitude of $Mg_{{}_D}$ should therefore be large enough in order for the kinetic effects to overcome this stabilizing convective effect.

\begin{figure}[htbp]
\includegraphics[width=210pt]{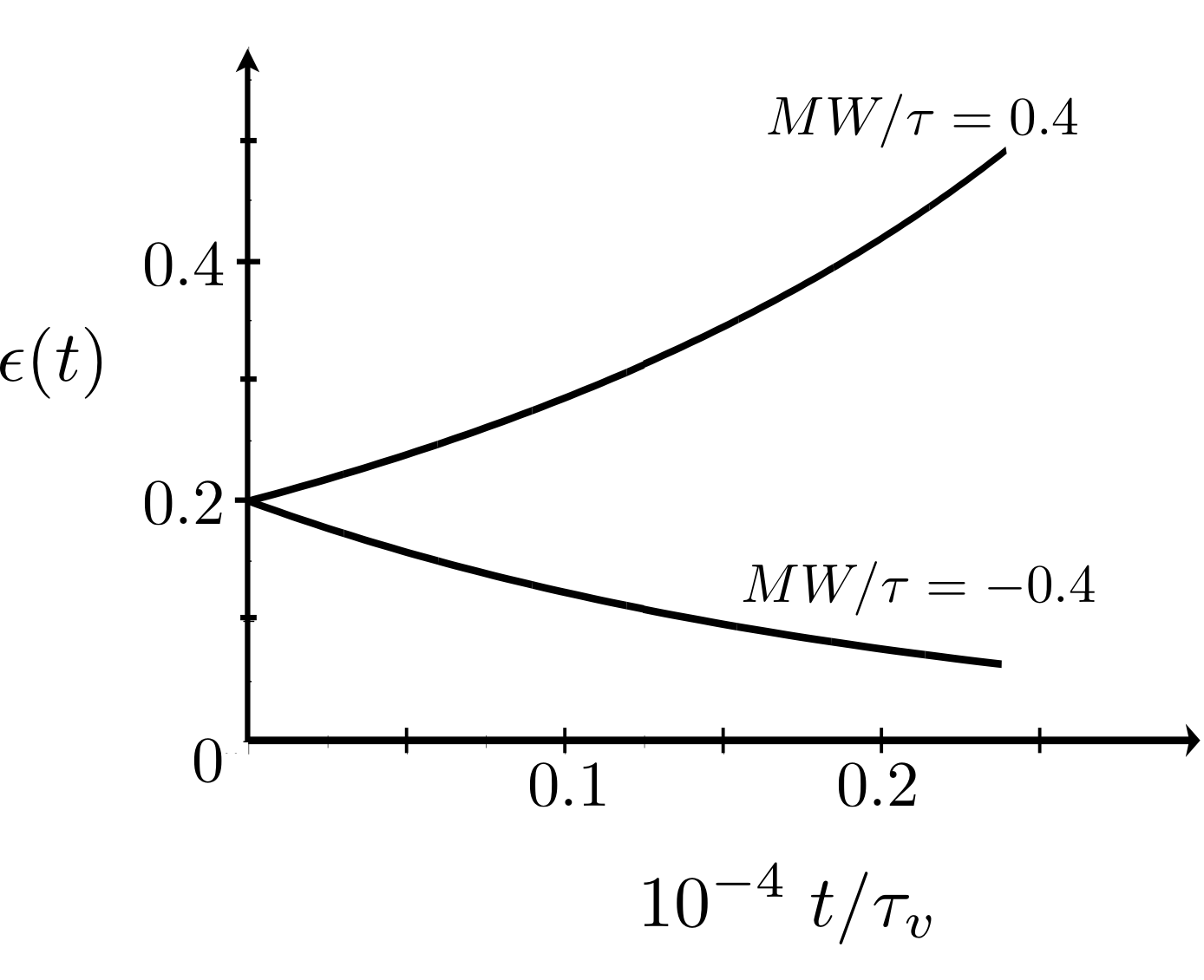}
\caption{\label{graph_inst} Time evolution of $\epsilon$ in the case $g_{{}_D}=5$ and for $MW/\tau=\pm0.4$. For $MW/\tau=0.4$, the vicinal surface is unstable, i.e. $\epsilon$ increases exponentially. For $MW/\tau=-0.4$, the vicinal surface is stable, i.e. $\epsilon$ decreases exponentially towards 0.}
\end{figure}

We made simulations with $F\tau_v - c_{eq}=0.025$, $L=20W$, $D\tau/W^2 = 20$ and $\tau_v/\tau=20$ leading to $\sigma=1$. At $t=0$, we set $\epsilon(t=0)=0.2$.
As an illustration of the influence on the stability of the vicinal surface of the ES effect ($\nu_+ \neq \nu_-$), we present, in Fig. \ref{graph_inst}, $\epsilon(t)$ for $g_{{}_D}=5$ and $MW/\tau = \pm 0.4$. For $MW/\tau=0.4$, the vicinal surface is unstable and $\epsilon(t)$ increases exponentially leading eventually to a collision of the paired steps for $\epsilon=1$. In opposition, for $MW/\tau=-0.4$, the vicinal surface is stable and $\epsilon(t)$ decreases exponentially towards $\epsilon=0$ and the step flow regime. 
Here, $F\tau_v - c_{eq}>0$ corresponds to the growth of the crystal, and the vicinal surface is unstable for $\nu_+<\nu_-$ ($M>0$). The opposite case where $\nu_+>\nu_-$ ($M<0$) is often considered \cite{pierre_louis, pimpinelli} to be more realistic, the instability therefore occurring for sublimation, i.e. $F\tau_v - c_{eq}<0$. We numerically checked however that changing simultaneously the 
sign of $F\tau_v - c_{eq}$ and $M$ leaves all observables unchanged up to the presently desired accuracy (it is not excluded that higher order calculations may exhibit odd powers of $M$ multiplied by the square of $F\tau_v - c_{eq}$).

For the comparisons of phase field simulations with the macroscopic approach, our aim is to focus on kinetic effects and the related instability due to the ES effect. For that purpose, we define 
%and in order to focus on the ES effect and the related instability, we define
\be
\tilde \lambda(M,g_{ {}_D}) = \lambda(M,g_{ {}_D}) - \lambda(-M,g_{ {}_D}) \;.
\ee
According to Eq. (\ref{lambda_pf}), $\tilde \lambda$ represents the first term on the r-h-s in Eq. (\ref{lambda_pf}) that vanishes for $M=0$ or $g_{ {}_D}=0$. We then compare $\lambda$ and $\tilde \lambda$ resulting from the phase field simulations to the semi-analytical ones computed numerically using the procedure presented in Appendix \ref{macro_steps} as mentioned above.
% The values of $D$ and $\tau_v$ used for the simulations do not allow us to compare quantitatively our numerical results to analytics using Eq. (\ref{lambda_pf}) since we have $\sqrt{D/\tau_v} \sim \nu_\pm$ and $\sqrt{D/\tau_v} \sim \nu_0$. Therefore we computed the theoretical $\tilde \lambda$ numerically using the procedure presented in the Appendix \ref{macro_steps}. 

\begin{figure}[htbp]
\includegraphics[width=230pt]{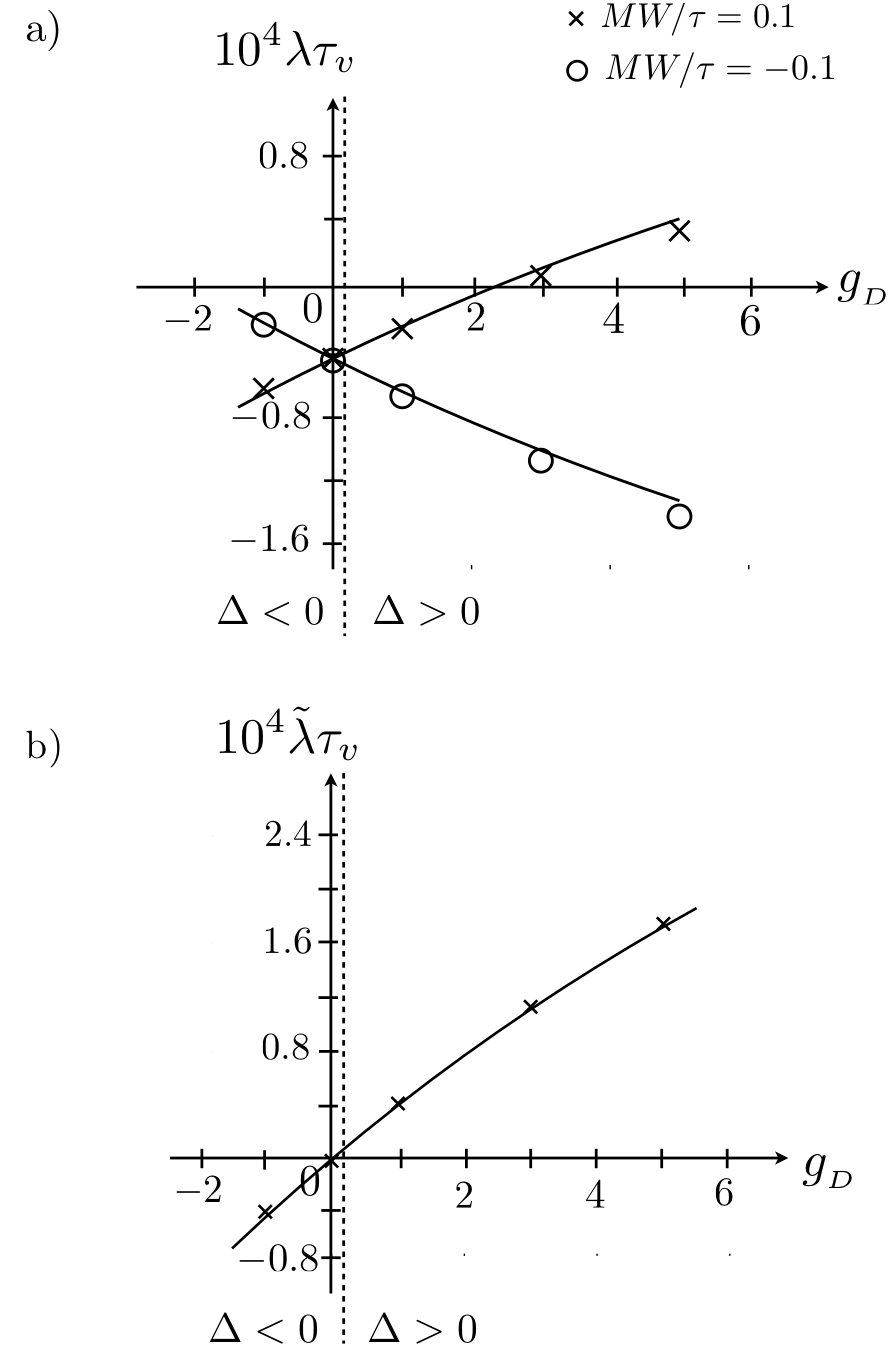}
\caption{\label{g_dependence} a) Dimensionless rate $\lambda \tau_v$ resulting from phase field simulations for $M\tau/W = 0.1$ (crosses) and $M\tau/W = -0.1$ (circles) and corresponding dimensionless rate within the macroscopic approach (line) as a function of $g_{ {}_D}$; b) Dimensionless rate $\tilde \lambda$ resulting from phase field simulations for $M\tau/W = 0.1$ (crosses) and corresponding dimensionless rate within the macroscopic approach (line) as a function of $g_{ {}_D}$.
In a) and b), the vertical dashed line separates the regions where the matrix of kinetic coefficients within the macroscopic approach is positive definite ($\Delta>0$) and where it is not ($\Delta<0$).}
\end{figure}

We first present the investigation of the dependence on $g_{ {}_D}$ of $\lambda$ and $\tilde \lambda$ resulting from phase field simulations together with the corresponding result within the macroscopic approach. We fix $|MW/\tau|=0.1$ and vary $g_{ {}_D}=-1,0,1,3,5$. In Fig. \ref{g_dependence}a), we present $\lambda \tau_v$ for $MW/\tau = \pm 0.1$, and in Fig. \ref{g_dependence}b), we present $\tilde \lambda \tau_v$ only for $MW/\tau = 0.1$ since by definition $\tilde \lambda(-M,g_{{}_D}) = - \tilde \lambda(M,g_{{}_D})$. In both cases, the simulations results are in good quantitative agreement with the macroscopic approach. 
Moreover, as expected qualitatively from Eq. (\ref{lambda_pf}), the instability ($\lambda >0$) occurs for sufficiently large $g_{{}_D}$ when $M>0$, and $\tilde \lambda$ vanishes when $g_{{}_D}=0$.
%
%the results for $\tilde \lambda$ compared with their analytical prediction computed numerically. $\tilde \lambda$ is normalized by the analytical prediction for $g_{ {}_D}=1$. The numerical results are in good quantitative agreement with the theory. The dependence on $g_{ {}_D}$ of $\tilde \lambda$ is not completely linear in opposition to the analytic formula Eq. (\ref{lambda_pf}). We have clear evidence that when the diffusional  resistance of the steps vanishes, i.e. $g_{{}_D}=0$, the rate of instability $\tilde \lambda$ vanishes also. Note also that in the case $g_{{}_D} = 3$ and $MW/\tau=0.1$, the convective effects [represented by the second term on the r-h-s of Eq. (\ref{lambda_pf})] compensate the destabilizing kinetic effects [represented by the first term on the r-h-s of Eq. (\ref{lambda_pf})]. This means that for $g_{{}_D}<3$, the vicinal surface is stable in both cases, i.e. $MW/\tau =0.1$ and $MW/\tau=-0.1$, for our choice of $F\tau_v - c_{eq}=0.025$. For $g_{{}_D}=5$, we show in Fig. \ref{graph_inst} that the vicinal surface is then unstable in the case $MW/\tau =0.1$ and stable in the case $MW/\tau=-0.1$
For $g_{ {}_D} > 0.2$, the determinant $\Delta$ of the macroscopic Onsager matrix in Eq. (\ref{delta}) is positive. In those cases, one may use the kinetic coefficients given by Eqs. (\ref{nupm}) and (\ref{nuo}) to perform time dependent calculations within the macroscopic approach and compare them to phase field simulations or to the semi-analytical solution computed numerically (lines in Fig. \ref{g_dependence}). As mentioned in the paragraph \ref{reduction_alloy_positiveness}, this is not possible in the opposite case ($\Delta <0$). This however does not prohibit semi-analytical calculations with $\Delta<0$ and their comparison with phase field simulations, as presented in the region to the left of the vertical dashed line in Fig. \ref{g_dependence}. 
%As also mentioned in the paragraph \ref{reduction_alloy_positiveness}, "exotic" materials may exhibit a matrix of kinetic coefficients that is not positive definite, i.e. that corresponds to $\Delta <0$. 
If $g_{{}_D}<0$, the stability of the vicinal surface is reversed and $\lambda >0$ with $F\tau_v-c_{eq}>0$ and $M<0$ (for sufficiently small $F\tau_v - c_{eq}$ in order to avoid convective effects). The vicinal surface is then unstable upon growth for a usual sign of the ES effect ($M<0$).

\begin{figure}[htbp]
\includegraphics[width=250pt]{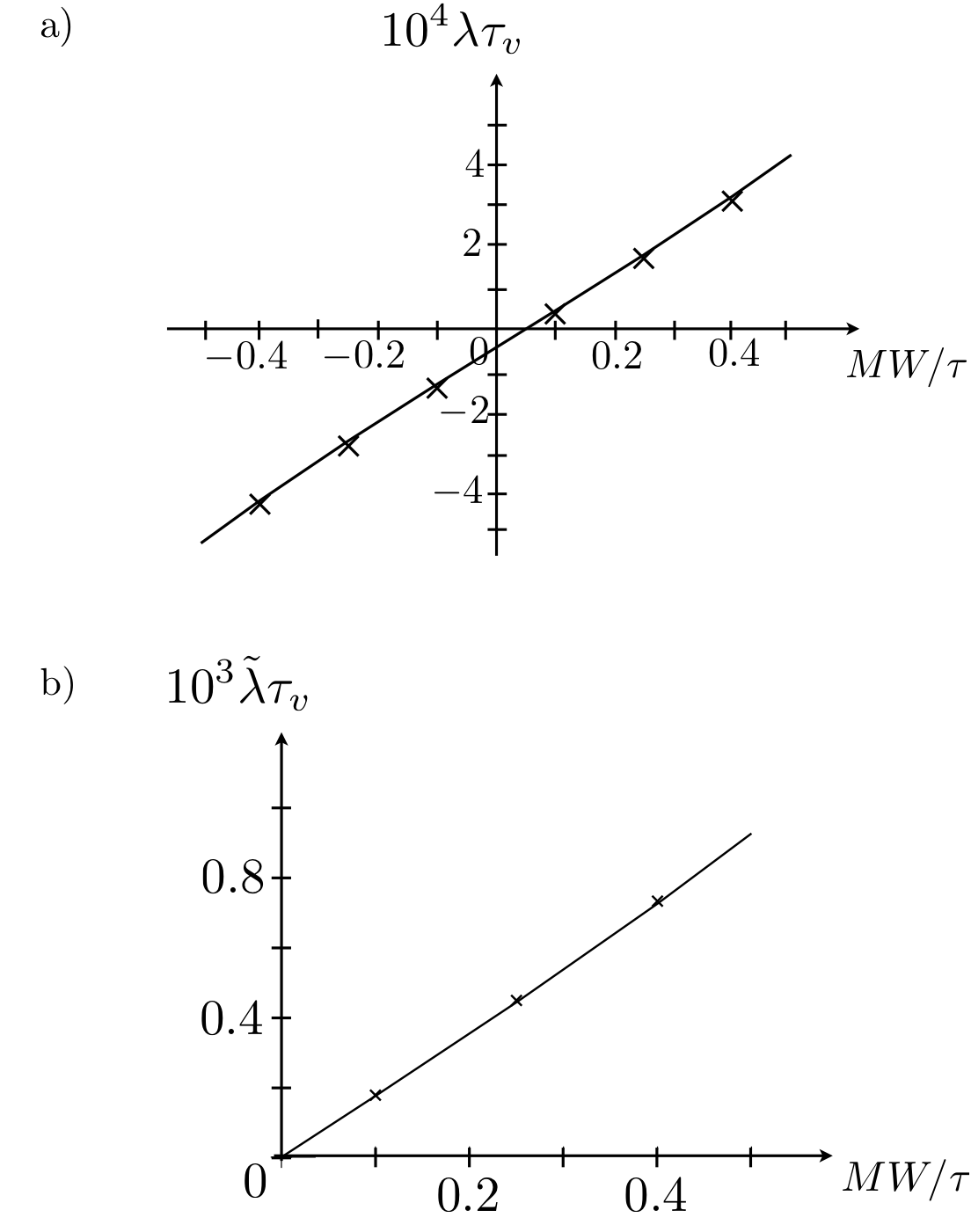}
\caption{\label{g_5_m_dependence} Comparison of phase field simulations (crosses) with the macroscopic approach (line) for $g_{ {}_D}=5$ for the dependence on $M$ of the dimensionless rate: a) $\lambda \tau_v$, b) $\tilde \lambda \tau_v$.  Here $\Delta>0$ for the presented range of $MW/\tau$ (see text for details).}
\end{figure}

Finally, we present, for $g_{ {}_D}=5$, the dependence on $M$ of $\lambda$ in Fig. \ref{g_5_m_dependence}a) and $\tilde \lambda$ in Fig. \ref{g_5_m_dependence}b). Again we present $\tilde \lambda$ only for positive values of $MW/\tau$ since $\tilde \lambda$ is an odd function by definition. 
Again, the phase field simulation results are in good quantitative agreement with the macroscopic approach. Moreover, as expected qualitatively from Eq. (\ref{lambda_pf}), the instability occurs ($\lambda>0$) for sufficiently large $M>0$ and $\tilde \lambda$ vanishes when $M=0$. The maximum value of $MW/\tau$, set by the stability condition (\ref{stability_mbe}), is 0.57. Moreover, for $MW/\tau < 0.51$, the determinant $\Delta$ is positive, and therefore all our calculations lie in this region.
%slightly deviates from linearity when $MW/\tau$ approaches its maximum value $0.57$ [set by the stability condition (\ref{stability_mbe})] and converges to 0 when $M \to 0$. For $MW/\tau < 0.51$, the determinant $\Delta$ is positive, and therefore all our calculations lie in this region.

\section{Summary}

We have presented a unified description of interface kinetic effects in phase field models for isothermal phase transformations in binary alloys  and for steps dynamics in molecular-beam-epitaxy (MBE). The phase field equations of motion are written in a variational form and incorporate the kinetic cross coupling between the phase field and the concentration field, presented in a recent Rapid Communication \cite{brener_boussinot}. This cross coupling generalizes the phenomenology of kinetic effects and was omitted in classical phase field models based on the model C within Hohenberg-Halperin classification \cite{hohenberg}. It corresponds to the terms parametrized by $M$ in Eqs. (\ref{free_energy}-\ref{diff_flux}) for binary alloys, and in Eqs. (\ref{phase_field_mbe}-\ref{continuity_mbe}) for steps dynamics in MBE. 
The stability of the phase field models, i.e. the positiveness of the dissipation function in Eq. (\ref{dissipation_pf}), restricts the magnitude of $|M|$.

In addition, we present the link between the phase field model, where the interface (or the step) is smooth and of finite width, and the macroscopic approach, where the interface is a boundary and infinitely sharp. We give in Eqs. (\ref{bara}-\ref{barc}) general expressions for the three independent elements of the symmetric $2\times2$ matrix of kinetic coefficients, that are describing the kinetic boundary conditions within the macroscopic approach.
The derivation is done using a physically motivated reduction procedure involving the calculation of the effective dissipation that may be ascribed to the interface. The positive definiteness of the matrix of kinetic coefficients is not guaranteed. This means that the domain of stability of the phase field model is wider than the one of the macroscopic approach.
The reduction procedure is equivalent to the thin interface limit \cite{karma_rappel} but is technically simpler. We thus  recover the well-known results of the thin interface limit for binary alloys in concise terms. In addition we derive the kinetic boundary conditions for steps dynamics (also described by three kinetic coefficients) in MBE corresponding to our phase field model. The Ehrlich-Schwoebel (ES) effect turns out to be provided by the cross coupling in the equations of motion, and disappears if $M=0$. 

We study numerically the step-bunching instability. The ES effect ($M\neq 0$) is actually not the sole ingredient for the instability to occur and a "diffusional resistance" of the step (analogous to the Kapitza resistance in the thermal problem), parametrized by $g_{{}_D}$, is also required within our model. We compare the results of the phase field simulations with analytical calculations within the macroscopic approach. This comparison includes some range of phase field parameters where the corresponding matrix of macroscopic kinetic coefficients is not positive definite.

\section*{ACKNOWLEDGMENTS.} We acknowledge the support of the Deutsche Forschungsgemeinschaft under Project SFB 917.

\appendix

\section{Solution for the one-dimensional dynamics of paired steps} \label{macro_steps}

In this appendix we present the one-dimensional macroscopic description (coordinate $x$) of paired steps on a vicinal surface. We describe the perturbation of the step-flow regime that leads to a difference of velocity of the two steps $V_1-V_2$.  

\subsection{Step-flow regime} 
In the step-flow regime, the steps are equidistant and move with the same velocity $V_1=V_2=V$. All terraces are thus equivalent with a length $L$. When the driving force is small $F\tau_v - c_{eq} \ll 1$,  the concentration field on a terrace obeys the diffusion equation in the static approximation:
\be
D c''(x) +F- \frac{c(x)}{\tau_v} = 0 \;, \nonumber
\ee
whose general solution reads
\be
c(x) = F\tau_v + A \exp(-x/l_v) + B \exp(x/l_v) \nonumber
\ee
where $l_v= \sqrt{D\tau_v}$. At $x=0$, the concentration is 
\be
c^+ =c(x=0) = F\tau_v + A + B \;. \nonumber
\ee
At $x=L$, the concentration is 
\be
c^- = c(x=L) = F\tau_v + A \exp(-\sigma) + B \exp(\sigma) \;. \nonumber
\ee
with $\sigma= L/l_v$.
We find
\bea
A = \frac{(X^+-X_v) \exp(\sigma) - (X^- -X_v) }{2 \sinh(\sigma)}  \nonumber\\
B = \frac{(X^- -X_v) - (X^+-X_v) \exp(-\sigma) }{2 \sinh(\sigma)} \nonumber
\eea
where $X^+ = c^+-c_{eq}$, $X^- = c^--c_{eq}$ and $X_v = F\tau_v- c_{eq}$.

One defines the fluxes
\bea 
J^+ &=& Dc'(x=0) = (B - A) J_v \nonumber\\
&=& \frac{(X^- -X_v) - (X^+ -X_v) \cosh(\sigma) }{ \sinh(\sigma)} J_v \nonumber \\
J^- &=& -Dc'(x=L) = [A \exp(-\sigma) - B \exp(\sigma)]  J_v\nonumber \\
&=& \frac{(X^+ -X_v) - (X^- -X_v) \cosh(\sigma) }{ \sinh(\sigma)} J_v \nonumber
\eea
where $J_v = D/l_v$.
The kinetic coefficients are defined such that
\bea
J^\pm = (\nu_\pm + \nu_0) X^\pm  - \nu_0 X^\mp \;. \nonumber
\eea
The velocity is then
\be
V = J^+ + J^- = \nu_+ X^+ + \nu_- X^-\;, \nonumber
\ee
and one finds
\bea
V &=& V_{eq} \; \frac{1+ J_v \Delta \frac{\nu_+ + \nu_-}{2} \frac{\cosh(\sigma) + 1}{\sinh(\sigma)} }{1+ 2 J_v \Delta \nu_0 \frac{\cosh(\sigma) - 1}{\sinh(\sigma)} + J_v \Delta \frac{\nu_+ + \nu_-}{\tanh(\sigma)} + J_v^2 \Delta }, \nonumber
\eea
where
\be
V_{eq} = 2 J_v X_v \frac{\cosh(\sigma) - 1}{\sinh(\sigma)} \nonumber
\ee
is the steady-state velocity of the step-flow regime when kinetic effects are absent ($\nu's \to \infty$), and where $\Delta = [\nu_+ \nu_- + \nu_0 (\nu_+ + \nu_-)]^{-1}$.  
%Let us note that in the limit $\sigma \ll 1$, we have $V = V_{eq} [1- (D/L) \Delta (\nu_+ + \nu_-)/2 - (D/\tau_v) \Delta ]$. In this limit, neglecting kinetic effects therefore requires $\sqrt{D/\tau_v} \ll \nu$'s and $D/L \ll \nu$'s. 

\subsection{Paired steps}

We now discuss the perturbation of the step-flow regime. The terraces are denoted by the integer $k$ and the $k$-th terrace has a length $L_k$. In the limit of small driving force $X_v \ll 1$, one has $V \ll J_v$, and the steady-state velocity $V$ enters the diffusion equation in the quasi-steady approximation for the concentration field $c_k(x)$ on the $k$-th terrace in the form:
%We separate the study of the step-bunching instability within the static approximation $\dot c=0$, which is of purely kinetic nature, and the study of the relaxation to step flow regime for equilibrium boundary conditions due to the correction to the concentration field when $\dot c \neq 0$. The two contributions for the difference of velocity of the two steps $V_1-V_2$ should then be added to account for both effects.
\be
D c_k''(x) +V c_k'(x)+F- \frac{c_k(x)}{\tau_v} = 0 \;. \nonumber
\ee
The solution reads 
\bea
c_k(x) = F \tau_v + A_k \exp \left( -x/l_D - x\sqrt{1/l_D^2 + 1/l_v^2 }  \right) \nonumber\\
+ B_k \exp \left( -x/l_D + x\sqrt{1/l_D^2 + 1/l_v^2 } \right) \nonumber\\
\simeq F \tau_v +  \exp(-x/l_D) \left[ A_k \exp(-x/l_v) + B_k \exp(x/l_v) \right] \nonumber
\eea
with $l_D=2D/V \gg l_v$.

At the $k$-th steps, corresponding for the $k$-th terrace to $x=0$, the concentration is 
\be
c^+_k =c_k(x=0) = F \tau_v + A_k + B_k \;. \nonumber
\ee
At the $(k+1)$-th step, corresponding for the $k$-th terrace to $x=L_k$, the concentration is 
\bea
c^-_k &=& c_k(x=L_k) \nonumber\\
&=& F \tau_v + \exp(-\beta \sigma_k) [A_k \exp(-\sigma_k) + B_k \exp(\sigma_k)] \;. \nonumber
\eea
with $\sigma_k = L_k/l_v$ and $\beta=l_v/l_D=V/(2J_v) \ll 1$.

One  finds
\bea
A_k = \frac{(X^+_k-X_v) \exp(\sigma_k) - (X^-_k -X_v)\exp(\beta \sigma_k) }{2 \sinh(\sigma_k)} \nonumber\\
B_k = -\frac{(X^+_k-X_v) \exp(-\sigma_k)-(X^-_k -X_v) \exp(\beta \sigma_k)  }{2 \sinh(\sigma_k)} \nonumber
\eea
where $X^+_k = c^+_k-c_{eq}$, $X^-_k = c^-_k-c_{eq}$.

One defines the fluxes
\bea
J^+_k &=& Dc_k'(x=0) = (-A_k+B_k)J_v - (A_k+B_k)V/2   \nonumber \\
&=& \frac{(X^-_k -X_v) \exp(\beta \sigma_k)- (X^+_k -X_v) \cosh(\sigma_k) }{ \sinh(\sigma_k)} J_v \nonumber \\ 
&& - (X^+_k-X_v)V/2 \nonumber
\eea
\bea
J^-_k &=& -Dc_k'(x=l_k) \nonumber \\
&=& [A_k \exp(-\sigma_k) - B_k \exp(\sigma_k)]  \exp(-\beta \sigma_k) J_v \nonumber \\
&& + [A_k \exp(-\sigma_k) + B_k \exp(\sigma_k)]  \exp(-\beta \sigma_k) V/2\nonumber \\
&=& \frac{(X^+_k -X_v) \exp(-\beta \sigma_k) - (X^-_k -X_v) \cosh(\sigma_k) }{ \sinh(\sigma_k)} J_v \nonumber \\
&& + (X^-_k -X_v) V/2 \nonumber
\eea

For $\beta \sigma_k \ll 1$, one therefore has
\bea
\frac{J_k^+ }{J_v} &\simeq& \frac{(X^-_k -X_v) - (X^+_k -X_v) \cosh(\sigma_k) }{ \sinh(\sigma_k)}  \nonumber \\ 
&& - \beta[ (X^+_k-X_v) - \sigma_k (X^-_k -X_v)/\sinh(\sigma_k) ] \nonumber \\ \label{J_k+}
\eea
\bea
\frac{J_k^-}{J_v} &\simeq& \frac{(X^+_k -X_v) - (X^-_k -X_v) \cosh(\sigma_k) }{ \sinh(\sigma_k)}  \nonumber \\
&& + \beta [ (X^-_k -X_v) - \sigma_k (X^+_k -X_v)/\sinh(\sigma_k) ] \nonumber \\ \label{J_k-}
\eea

Since the kinetic coefficients are defined by
\be
J_k^\pm = (\nu_\pm + \nu_0) X_k^\pm - \nu_0 X_{k\mp1}^{\mp} , \nonumber
\ee
we have the set of coupled equations
\be
\mathcal N^{\pm}_k X_k^\pm =  \mathcal I^\pm_k X_k^\mp + \nu_0 X_{k\mp 1}^{\mp} + \mathcal M_k^\pm X_v  \nonumber
\ee
with 
\bea
\mathcal N^{\pm}_k &=& \nu_\pm + \nu_0 + J_v\big( 1/\tanh (\sigma_k) \pm \beta \big) \nonumber \\
\mathcal I^\pm_k &=& J_v (1\pm \beta \sigma_k ) / \sinh (\sigma_k) \nonumber \\
 \mathcal M_k^\pm &=& J_v \left[ \frac{ \cosh(\sigma_k) - 1 }{\sinh (\sigma_k)} \pm \beta  \big(1-\sigma_k/\sinh(\sigma_k) \big) \right] \;. \nonumber
\eea
The velocity of the $k$-th step is then
\be
V_k = \nu_+ X_k^+ + \nu_- X_{k-1}^-. \nonumber
\ee

For paired steps one has $L_k = L [1+(-1)^k \epsilon]$, and then $\sigma_k =L_k/l_v= \sigma + (-1)^k \delta \sigma$ where $\sigma=L/l_v$ and $\delta \sigma = \epsilon \sigma$, such that $\sigma_1 = \sigma - \delta \sigma$ and $\sigma_2 = \sigma + \delta \sigma$.
The system of equations reads
\be
\mathcal N^{\pm}_i X_i^\pm =  \mathcal I^\pm_i X_i^\mp + \nu_0 X_{j}^{\mp} + \mathcal M_i^\pm X_v \nonumber
\ee
where $(i,j) = (1,2)$ or $(i,j) = (2,1)$.

The solution is then given by
\be
X_i^\pm = \frac{\chi_i^\pm \gamma_j^\pm + \rho_j^\pm \gamma_i^\pm}{\rho_i^\pm \rho_j^\pm - \chi_i^\pm \chi_j^\pm} X_v \nonumber
\ee
with
\bea
\chi_i^\pm &=& \nu_0(\m I_i^\pm \m N_j^\mp + \m I_j^\mp \m N_i^\mp ) \nonumber \\
\rho_i^\pm &=& ( \m N_i^+ \m N_i^-  - \m I_i^+ \m I_i^-) \m N_j^\mp  - \nu_0^2 \m N_i^\mp \nonumber \\
\gamma_i^\pm &=& \m N_i^\mp (\nu_0 \m M_j^\mp + \m N_j^\mp \m M_i^\pm) + \m N_j^\mp \m I_i^\pm \m M_i^\mp . \nonumber
\eea

The difference of velocity is then 
\be
V_1-V_2 = L \dot \epsilon = \nu_+ (X_1^+ - X_2^+) - \nu_- (X_1^- - X_2^-) \;. \nonumber
\ee
The rate of growth or decay  $\lambda = \dot \epsilon/\epsilon = (V_1-V_2)/(L\epsilon)$ of the perturbation was computed numerically in order to be compared with the phase field simulation results. 

However, some analytical progress may be made when, in addition to the assumption $X_v \ll 1$, one assumes that kinetic effects are small, i.e. that the $\nu's$ are much larger than the two velocity scales $\sqrt{D/\tau_v}$ and $D/L$. In this case, $\lambda$ contains two main contributions. 

{\it Kinetic effects in the static approximation.} The first contribution to $\lambda$ is due to the Ehrlich-Schwoebel effect and is present in the static approximation $\beta = 0$. It is proportional to the driving force $X_v$ and it may be shown by straightforward but tedious algebra that it corresponds to the first term on the r-h-s in Eq. (\ref{lambda}). 

{\it Relaxation to the step-flow regime with equilibrium boundary conditions.} The second contribution arises from the convection effect $\beta \neq 0$, is proportional to $X_v^2$ and is present without kinetic effects. Setting $X_k^\pm = 0$ and $V=V_{eq}$ in Eqs. (\ref{J_k+}) and (\ref{J_k-}), one obtains for paired steps with $J_k^\pm = J_{k+2}^\pm$:
\be
V_1 - V_2 = V_{eq} X_v \big( \sigma_2 / \sinh(\sigma_2) - \sigma_1 / \sinh(\sigma_1)  \big) \nonumber.
\ee
In the limit $\epsilon \ll 1$, one obtains the rate 
\be
\dot \epsilon/\epsilon = 2 X_v \frac{V_{eq}}{l_v} \; \frac{\sinh(\sigma) - \sigma \cosh(\sigma)}{\sinh^2(\sigma)}
\ee
that corresponds to the second term on the r-h-s in Eq. (\ref{lambda}). This term is negative and therefore promotes a relaxation to the step-flow regime and the stability of the vicinal surface.

\section{Phase field equations of motion using an Onsager matrix giving fluxes in terms of driving forces} \label{other_pf}

One may write the equations of motion for the phase field $\phi$ and the concentration field $C$ using an Onsager matrix that gives fluxes in terms of driving forces, a representation that may be more familiar to the reader. The equations of motion then read
\bea
\dot \phi &=&  (-\delta G/\delta \phi)/\mathcal T(\phi) + [\mathcal J(\phi) W\boldnabla \phi] \cdot (-\boldnabla \delta G/\delta C) \nonumber\\
{\bf J} &=& [\mathcal J(\phi) W\boldnabla \phi] (-\delta G/\delta \phi) + \mathcal D(\phi) (-\boldnabla \delta G/\delta C) \nonumber\\
\dot C &=& - \boldnabla \cdot {\bf J} , \nonumber
\eea 
with the interface width $W$.

The positive definiteness of the Onsager matrix requires $\mathcal T(\phi)>0$, $\mathcal D(\phi) >0$ and $ \mathcal D(\phi) /\mathcal T(\phi)> [\mathcal J(\phi) W \boldnabla \phi]^2$.
The time derivative of the phase field and the continuity equation are then written in a simple variational form: 
\bea
\dot \phi &=&  - \frac{1}{\mathcal T(\phi)} \frac{\delta G}{\delta \phi}  - \mathcal J(\phi) W\boldnabla \phi \cdot \boldnabla \frac{\delta G}{\delta C} \label{phi_dot}\\
\dot C &=&  \boldnabla \cdot \left( \mathcal D(\phi) \boldnabla \frac{\delta G}{\delta C} \right)+ \boldnabla \cdot \left(\mathcal J(\phi) W\frac{\delta G}{\delta \phi} \boldnabla \phi\right) \;. \label{c_dot}
\eea
When $\mathcal J=0$, one recovers the diagonal model, i.e. model C \cite{hohenberg}. For $\mathcal J \neq 0$, non diagonal terms are present providing a third kinetic velocity scale.

The link with the parameters $\tau(\phi)$, $D(\phi)$ and $M(\phi)$ that enter Eqs. (\ref{phase_field}) and (\ref{diff_flux}) where the driving forces are given in terms of the fluxes is provided by:
\bea
\mathcal T (\phi) &=& \tau(\phi) \Delta_{PF} \nonumber \\
\mathcal D(\phi) &=& \frac{D(\phi)}{\Delta_{PF}} \nonumber \\
\mathcal J(\phi) &=& - \frac{M(\phi) D(\phi)}{\tau(\phi) \Delta_{PF}}   \label{link}
\eea
where the determinant given in Eq. (\ref{determinant}):
\be
\Delta_{PF} = 1 - \frac{M^2(\phi) D(\phi) (W \boldnabla \phi)^2}{\tau(\phi)} = 1- \frac{\mathcal T(\phi) \mathcal J^2(\phi) (W \boldnabla \phi)^2}{\mathcal D(\phi)}
\ee
is independent of the used representation. The kinetic coefficients $\bar \ma, \bar \mb$ and $\bar \mc$ may then be obtained in terms of $\mathcal T$, $\mathcal J$ and $\mathcal D$ inserting the relations (\ref{link}) in Eqs. (\ref{bara}), (\ref{barb}) and (\ref{barc}). 

In sections \ref{examples_alloy} and \ref{model_mbe}, we assume for simplicity that $M$ and $\tau$ are constants. In this frame, the coefficients $\mathcal T$ and $\mathcal J$ are thus $\phi$-dependent. However, using the representation where fluxes are given in terms of driving forces, one may as well assume that $\mathcal T$ and $\mathcal J$ are constants, a choice that simplifies the implementation of the equations of motion Eqs. (\ref{phi_dot}) and (\ref{c_dot}). 

%The kinetic coefficient $\bar \ma, \bar \mb$ and $\bar \mc$ may be derived in the same way as presented in section \ref{alloy} with the total dissipation in the system written as:
%\bea
%R &=& \frac{1}{2} \int_V dV \left( - \dot \phi \frac{\delta G}{\delta \phi} - {\bf J} \cdot \boldnabla \frac{\delta G}{\delta C} \right) \nonumber\\
%&=& \; \frac{1}{2} \int_V dV \frac{\mathcal T \big(\dot \phi \big)^2 + {\bf J}^2/\mathcal D(\phi) - 2 \mathcal T \mathcal JW \dot \phi \boldnabla \phi \cdot {\bf J}/\mathcal D(\phi)}{1- \mathcal T (\mathcal JW\boldnabla \phi)^2/\mathcal D(\phi)} \nonumber \;.
%\eea

%This leads to
%\bea
%\bar \ma &=& \mathcal T \int_{-\infty}^\infty dx \frac{[\phi_{eq}'(x)]^2 \mathcal D(\phi_{eq})}{\tilde D(\phi_{eq})} \nonumber\\
%&&+ 2 \mathcal T \mathcal J W \int_{-\infty}^\infty dx \frac{[\phi_{eq}'(x)]^2 C_{eq}(x)}{\tilde D(\phi_{eq})} \nonumber \\
%&&+ \int_{-\infty}^\infty dx \left[ \frac{C^2_{eq}(x)}{\tilde D(\phi_{eq})} - \frac{(C_1^{eq})^2}{2D_1} - \frac{(C_0^{eq})^2}{2D_0} \right] \nonumber ,
%\eea

%\bea
%\bar \mb &=& - \mathcal T \mathcal J W \int_{-\infty}^\infty dx \frac{[\phi_{eq}'(x)]^2}{\tilde D(\phi_{eq})} \nonumber \\
% &&- \int_{-\infty}^\infty dx \left[ \frac{C_{eq}(x)}{\tilde D(\phi_{eq})} - \frac{C_1^{eq}}{2D_1} - \frac{C_0^{eq}}{2D_0} \right] \nonumber ,
%\eea

%\bea
%\bar \mc = \int_{-\infty}^\infty dx \left[ \frac{1}{\tilde D(\phi_{eq})} - \frac{1}{2D_1} - \frac{1}{2D_0} \right] \nonumber ,
%\eea
%where
%\be
%\tilde D(\phi_{eq}) = \mathcal D(\phi_{eq}) \Delta_{PF}=\mathcal D(\phi_{eq}) - \mathcal T \mathcal J^2 [W \phi_{eq}'(x)]^2 \;. \nonumber
%\ee


\begin{thebibliography}{99}
\bibitem{comment}Strictly, the parameter $D$ in Eq. (\ref{diff_flux}) refers to the diffusional mobility rather than to diffusion coefficient. In the bulk of phase $i$, the diffusional mobility is equal to the diffusion coefficient in phase $i$ divided by $g''_i(C_i^{eq})$, i.e. the second derivative with respect to the concentration $C$ of the dimensionless free energy $g_i(C)$ of phase $i$ (at the equilibrium concentration $C_i^{eq}$). For simplicity, we omit this difference in the rest of the article by choosing  bulk free energies of appropriate parabolic form leading to $g''_i(C_i^{eq})=1$.
\end{thebibliography}
\end{document}